\documentclass[aps,reprint,superscriptaddress,twocolumn,floatfix,showpacs,longbibliography]{revtex4-1}

\usepackage{graphicx,color}
\usepackage[utf8]{inputenc}
\usepackage{amsmath,amssymb,bm,amsfonts,dsfont,mathrsfs,amsthm}
\usepackage{braket}
\usepackage{txfonts,comment}
\usepackage[colorlinks=true,linkcolor=blue,citecolor=blue,urlcolor=blue]{hyperref}
\usepackage{soul}
\usepackage{enumitem}
\usepackage{multirow}
\usepackage{orcidlink}

\newcommand{\nn}{\nonumber \\}

\newcommand{\ketbra}[2]{\lvert#1\rangle\langle#2\rvert}
\newcommand{\norm}[1]{\lVert#1\rVert}
\newtheorem{theorem}{Theorem}[section]

\AtBeginDocument{%
    \newwrite\bibnotes
    \def\bibnotesext{Notes.bib}
    \immediate\openout\bibnotes=\jobname\bibnotesext
    \immediate\write\bibnotes{@CONTROL{REVTEX41Control}}
    \immediate\write\bibnotes{@CONTROL{%
    apsrev41Control,author="08",editor="1",pages="1",title="0",year="1"}}
     \if@filesw
     \immediate\write\@auxout{\string\citation{apsrev41Control}}%
    \fi
}%

\begin{document}

\title{Quantum-enhanced sensing in a driven-dissipative system via chiral waveguide}

\author{Yan Xi Foo\,\orcidlink{0009-0001-0811-4673}}
\email{fooy0044@e.ntu.edu.sg}
\affiliation{Quantum Science and Engineering Centre, Nanyang Technological University, Singapore 639798}
\affiliation{Centre for Quantum Technologies, National University of Singapore, Singapore 117543}

\author{Saubhik Sarkar\,\orcidlink{0000-0002-2933-2792}}
\email{saubhik.sarkar@uestc.edu.cn}
\affiliation{Institute of Fundamental and Frontier Sciences, University of Electronic Science and Technology of China, Chengdu 611731, China}
\affiliation{Key Laboratory of Quantum Physics and Photonic Quantum Information, Ministry of Education, University of Electronic Science and Technology of China, Chengdu 611731, China}

\author{Leong Chuan Kwek\,\orcidlink{0000-0002-0879-0591}}
\email{cqtklc@nus.edu.sg}
\affiliation{Centre for Quantum Technologies, National University of Singapore, Singapore 117543}
\affiliation{National Institute of Education, Nanyang Technological University, Singapore 637616}

\author{Abolfazl Bayat\,\orcidlink{0000-0003-3852-4558}}
\email{abolfazl.bayat@uestc.edu.cn}
\affiliation{Institute of Fundamental and Frontier Sciences, University of Electronic Science and Technology of China, Chengdu 611731, China}
\affiliation{Key Laboratory of Quantum Physics and Photonic Quantum Information, Ministry of Education, University of Electronic Science and Technology of China, Chengdu 611731, China}
\affiliation{Shimmer Center, Tianfu Jiangxi Laboratory, Chengdu 641419, China}

\author{Davit Aghamalyan\,\orcidlink{0009-0007-4926-9739}}
\email{davagham@gmail.com}
\affiliation{SMT Pillar, Singapore University of Technology and Design, Singapore 487372}

\begin{abstract}

Quantum sensors can achieve precision unattainable by classical sensors.
However, dissipation coming from interaction with the environment can greatly degrade the performance of quantum sensors.
Therefore, it is crucial to attempt harnessing dissipation to our advantage and investigate the sensing performance of driven-dissipative systems to gain practical quantum advantages.
In this paper, we study a system of two-level systems driven coherently at a detuned frequency, and importantly, connected to a chiral waveguide that serves both as an interaction mediator and a bath.
We show analytically that the steady state of such a system shows enhanced sensitivity to estimate weak detuning strength while the probe preparation time only grows linearly with system size.
Therefore, the enhanced sensitivity is sustained even when the preparation time is incorporated.
We further establish measurement protocols to achieve such sensitivity that are experimentally implementable.
The chirality of the waveguide helps us to control both precision and range of enhanced sensitivity. 

\end{abstract}

\maketitle

\section{Introduction} 

Quantum sensors utilize quantum features to accomplish sensitivity beyond the capability of classical sensors~\cite{degen2017quantum, braun2018quantum, ye2024essay, ghosh2026journey}.
The sensitivity can be quantified with the precision of estimating an unknown parameter in a $N$-size system that generally scales as $N^{\beta}$.
While the scaling exponent $\beta {=} 1$ denotes the standard quantum limit that can be achieved with classical protocols, true quantum sensing advantage is signaled by $\beta {>} 1$.
Several resources for such quantum advantages have been identified, such as entanglement in Greenberger-Horne-Zeilinger (GHZ) type states~\cite{giovannetti2004quantum, giovannetti2006quantum, giovannetti2011advances}, squeezing in optical~\cite{maccone2020squeezing, pezze2008mach, schnabel2017squeezed, polino2020photonic} or spin states~\cite{ma2011quantum, frerot2018quantum}, and quantum phase transitions~\cite{venuti2007quantum, zanardi2008quantum, rams2018limits, montenegro2025review}. 
These advantages can however seriously suffer from decoherence in practice as unavoidable interaction with the environment renders the sensing probe an open quantum system~\cite{demkowicz2012elusive, de2013quantum}.
Quantum phase transitions in the steady states of open quantum systems provide with a promising avenue to engineer probe states with driven dissipation that shows quantum enhancement~\cite{fernandez2017quantum, heugel2019quantum, garbe2020critical, montenegro2023quantum, beaulieu2025criticality, ding2022enhanced, liu2022deep, wang2026quantum, liu2026enhancedmultiparameter, zhang2024rydberg, liu2021experimental, puig2025dynamical, yu2026dose, zhang2025performance}.
However, the time needed to prepare such critical probe states often suffer from criticality slowing-down~\cite{macieszczak2016towards, garbe2020critical,zeng2026global}, like in closed quantum systems~\cite{sachdev1999quantum}.
Therefore, it is crucial to identify open quantum systems that offer favorable scaling exponents for sensing while the preparation time does not suffer from too costly scaling properties.

Another avenue for tackling the noisy quantum systems is given by considering evolution under an effective non-Hermitian Hamiltonian where enhanced sensitivity have been found both theoretically~\cite{Wiersig2014Enhancing, Lau2018Fundamental, Zhang2019Quantum, budich2020non, McDonald2020Exponentially, sarkar2024critical} and experimentally~\cite{Liu2016Metrology, Hodaei2017Enhanced, Yu2020Experimental, Wang2020Petermann, xiao2024non, Yu2024Heisenberg, xiao2026observation}.
Nonetheless, this approach does not fully correspond to an open-system evolution governed by a Liouvillian but gives rise to an important question: can the properties associated with the non-Hermiticity of the Liouvillian operator lead to generating probe states with enhanced sensitivity?

In this paper, we provide an affirmative answer to these issues by exploiting dark state engineering.
We consider a system of $N$ coherently driven two-level systems/emitters that are coupled to a chiral waveguide~\cite{lodahl2017chiral,ramos2014quantum, pichler2015quantum, gonzalez2015chiral, SurezForero2025, ramos2016non, guimond2016chiral, mok2020long, lim2024exponentially, mok2020microresonators, sollner2015deterministic, ostfeldt2022demand, sayrin2015nanophotonic, junge2013strong, petersen2014chiral, le2017nanofiber}, where the degree of chirality fundamentally alters how the emitters interact with the field and with one another.
Chirality in quantum optics manifests~\cite{lodahl2017chiral, Sheremet2023, SurezForero2025}, for example, in systems where atoms are coupled to nanophotonic waveguides~\cite{sollner2015deterministic,ostfeldt2022demand}, such that an emitted photon preferentially propagates in one direction rather than distributing symmetrically between left- and right-moving modes. 
It has been realized experimentally in nanophotonic and photonic-crystal waveguide platforms~\cite{sollner2015deterministic, ostfeldt2022demand, petersen2014chiral}, as well as with cold atoms coupled to whispering-gallery-mode microresonators~\cite{sayrin2015nanophotonic, junge2013strong, le2017nanofiber}. 
Such chiral light-matter interfaces have since been employed for engineering directional photon routing~\cite{mitsch2014quantum}, non-reciprocal devices~\cite{scheucher2016quantum, sayrin2015nanophotonic,Miao2025}, tailored many-body dynamics in waveguide quantum electrodynamics~\cite{Mahmoodian2018, Mahmoodian2020, Soro2022, Irfan2024, lim2024exponentially, Windt2025, Wu2026}, as well as metrological applications~\cite{Kleinbeck2023, Takada2023}.

In our coupled emitter-waveguide system, the steady-state serves as the sensing probe, and it can estimate detuning strength with enhanced precision. 
The detuning can come from the bare frequency difference between the two-level systems and the drive or due to the presence of a field, enabling the system to act as a frequency sensor or a field sensor, respectively.
We show that the lower bound on the sensitivity scales as $N^3$ for uniform detuning and as $N^5$ for detuning with a linear gradient.
Importantly, the preparation time only scales linearly with $N$, which establishes a clear advantage even when the time is taken into consideration.
Both of the scaling properties follow from the structure of the Liouvillian operator, as we analytically show in the paper.
We then provide with measurement protocols that can be experimentally implemented on fairly small system sizes and still produce relatively high precision. 
Furthermore, the degree of chirality is shown to play a crucial role in controlling both the sensitivity and the range of enhanced-sensing capability.

\section{Overview of quantum sensing}
\label{sec:sensing}

We start with an overview of the single parameter estimation problem where an unknown parameter $\theta$ is encoded onto a quantum probe state denoted by $\rho_\theta$~\cite{paris2009quantum}.
Measurements are then performed on the probe state and estimation of $\theta$ is inferred from the outcomes with the aid of an estimator function.
The measurement can be generically described by a complete set of positive operator-valued measurement (POVM) operators $\{\Pi_m\}$ where the probability of the $m$-th outcome is given by $p_m(\theta) {=} \text{Tr}\left[\rho_\theta \Pi_m\right]$.
This classical probability leads to a statistical lower bound for the accuracy of estimation, given by the variance $\sigma_\theta^2$, in terms of the Cram\'er-Rao inequality $\sigma_\theta^2 \ge 1/\mathcal{M} F^C$ for unbiased estimators.
Here, the total number of measurements is $\mathcal{M}$ and the classical Fisher information (CFI), 
\begin{align}
    F^C = \sum_m p_m (\partial_\theta\ln p_m)^2
\end{align}
is dependent on the measurement basis~\cite{paris2009quantum}. 
A basis-independent way of determining the sensitivity is given by the quantum Fisher information (QFI) $F^Q$, which is the maximum of the CFI over all possible choices of basis and therefore, gives the ultimate lower bound.
One can then define a quantum Cram\'er-Rao bound
\begin{align}
\sigma_\theta^2 \ge \frac{1}{\mathcal{M} F^C} \ge \frac{1}{\mathcal{M} F^Q} \,.
\label{eq:cramer-rao}
\end{align}
The eigen-decomposition $\rho_{\theta} {=} \sum_{n, (\lambda_n \ne 0)} \lambda_n \ket{\lambda_n} \bra{\lambda_n}$ can be used to express the QFI in the following way~\cite{liu2019quantum},
\begin{align}
F^Q &= \sum_{n, (\lambda_n \ne 0)} \frac{(\partial_{\theta} \lambda_n)^2}{\lambda_n} 
+ \sum_{n, (\lambda_n \ne 0)} 4 \lambda_n \text{Re}(\braket{\partial_{\theta} \lambda_n | \partial_{\theta} \lambda_n}) \nn 
&- \sum_{\substack{n,m \\ (\lambda_n, \lambda_m \ne 0)}} 8 \frac{\lambda_n \lambda_m}{\lambda_n+\lambda_m} \text{Re}(\braket{\partial_{\theta} \lambda_n | \lambda_m} \braket{\lambda_m | \partial_{\theta} \lambda_n}) .
\label{eq:QFI_mixed}
\end{align}
For pure states  $\rho_{\theta} {=} \ket{\psi_{\theta}} \bra{\psi_{\theta}}$, one can simplify the expression $F^Q = 4\left(\braket{\partial_\theta \psi_{\theta}|\partial_\theta \psi_{\theta}} - |\braket{\partial_\theta \psi_{\theta}|\psi_{\theta}}|^2 \right)$.
The ultimate precision limit on the estimation is given by the QFI and the optimal basis for obtaining this bound is not unique.
However, one choice is always given by the projectors formed from the eigenvectors of the symmetric logarithmic derivative (SLD) operator $\mathscr{L}$, implicitly defined as $\partial_{\theta}\rho_{\theta} {=} (\rho_\theta \mathscr{L}_\theta {+} \mathscr{L}_\theta \rho_\theta)/2$.
An experimentally relevant way of calculating the sensitivity dependent on an observable $\mathcal{O}$ is given by the error-propagation formula~\cite{sidhu2020geometric}
\begin{align}
\sigma_\theta^2 = \frac{\braket{\mathcal{O}^2} - \braket{\mathcal{O}}^2}{|\partial_{\theta} \braket{\mathcal{O}}|^2} \equiv  \frac{1}{\mathcal{F}^E} ,
\label{eq:error}
\end{align}
where we define $\mathcal{F}^E$ as an effective Fisher information like object that serves as a lower bound for CFI and QFI~\cite{pezze2018quantum}
\begin{align}
F^Q \ge F^C \ge \mathcal{F}^E .
\label{eq:FI}
\end{align}

\section{Model}
\label{sec:model}

\begin{figure}[t]
\centering
\includegraphics[width=0.49\textwidth]{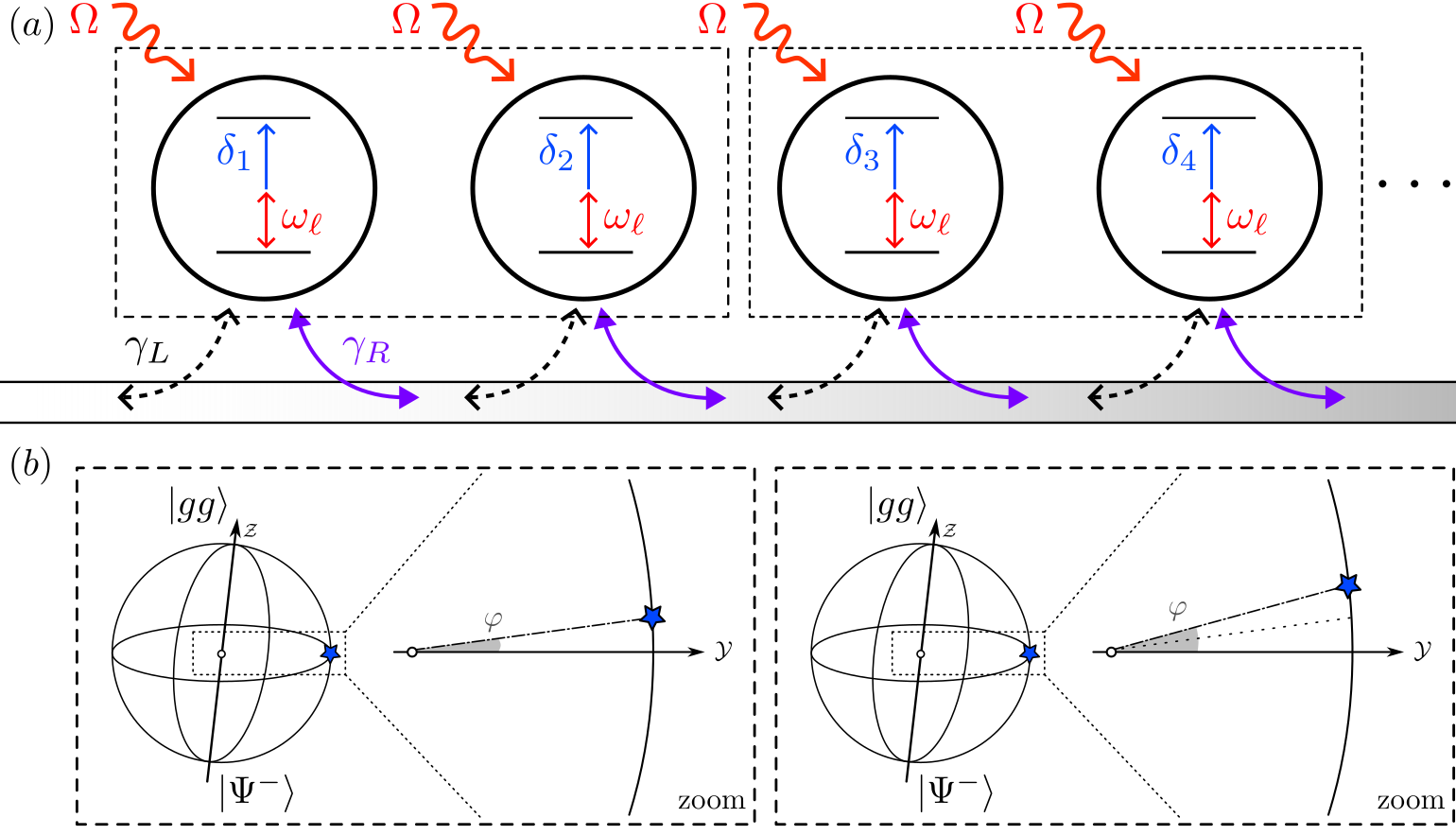}
\caption{\textbf{Schematic of the model}. 
(a) $N$ two-level systems or emitters are arranged in a 1D array and are connected to a 1D chiral waveguide.
The emitters are coherently driven with Rabi frequency $\Omega$ and driving frequency $\omega_\ell$.
The detunings $\delta_j$ are position dependent as the transition frequencies of the emitters $\omega_j$ can have $j$-dependence.
Due to the chirality of the waveguide, the photons are emitted with different rates $\gamma_L$ and $\gamma_R$ into the left and right propagating modes.
The waveguide mediates long-range interaction as the emitted photons can drive transition at a distance.
(b) The effect of a weak detuning strength can be perturbatively accounted for by the corresponding change for each dimer in a Bloch sphere in the dark state manifold.
Each successive dimer registers increasing change in the azimuthal phase $\varphi$, which can be used to calculate the lower bound of scaling of Fisher information with system size $N$.}
\label{fig:schematic}
\end{figure}

As shown in Fig.~\ref{fig:schematic}(a), we consider a 1D array of $N$ emitters with two levels that are labeled as $\ket{g}$ and $\ket{e}$.
The transition frequency of the $j$-th emitter is $\omega_j$ which one can control with an external field.
The emitters are coherently driven with Rabi frequency $\Omega$ and driving frequency $\omega_\ell$, giving rise to a detuning profile $\delta_j {=} \omega_\ell - \omega_j$.
The emitters are further coupled to a 1D waveguide which is chiral in nature as the decay rates for emitted photons into the left and right propagating modes, $\gamma_L$ and $\gamma_R$, are different in general.
The waveguide also mediates long-distance interaction between the emitters.
In the fully chiral limit, this realizes the cascaded open-system setting introduced in ~\cite{Carmichael1993, Gardiner1993} for which driven-dissipative dark-state stabilization was developed in ~\cite{Stannigel2012, pichler2015quantum}.
The Lindbladian master equation for the density operator $\rho$ for the emitters after tracing out the waveguide degrees of freedom were derived in Ref.~\cite{pichler2015quantum} under the standard rotating wave approximation and Born-Markov approximation and can be written down as 
\begin{align}
\dot{\rho} = \mathcal{L} [\rho] = -i [H, \rho] &+ \gamma_R \left( c_R \rho c_R^{\dagger} - \frac{1}{2} c_R^{\dagger} c_R \rho -  \frac{1}{2} \rho c_R^{\dagger} c_R  \right) \nn
&+ \gamma_L \left( c_L \rho c_L^{\dagger} - \frac{1}{2} c_L^{\dagger} c_L \rho - \frac{1}{2} \rho c_L^{\dagger} c_L  \right) ,
\label{eq:master}
\end{align}
where $\mathcal{L} [\rho]$ is the Liouvillian superoperator, $H$ is the system Hamiltonian
\begin{equation}
\begin{split}
 H {=} &\sum_{j=1}^N\left(-\delta_j\sigma_j^+\sigma_j^- {+} \frac{\Omega}{2}\sigma_j^- {+} \frac{\Omega^*}{2}\sigma_j^+\right)\\ 
 &{+} \sum_{j<k}^{ }\left[\frac{i}{2}\left(\gamma_R e^{-i\phi_{jk}} {-} \gamma_L e^{+i\phi_{jk}} \right)\sigma_j^+\sigma_k^- {+}\mathrm{H.c.}\right] ,
\label{eq:ham}
\end{split}
\end{equation}
and $c_L {=} \sum_j e^{i\phi_{j}}\sigma_j^-$ and $c_R {=} \sum_j e^{-i\phi_{j}}\sigma_j^-$ are the collective jump operators with $\sigma_j^- {=} \ketbra{g}{e}_j$ as the lowering operator for the $j$-th emitter. 
While $\phi_j$ is the phase associated with the travelling modes, $\phi_{jk}$ is the phase accumulated due to waveguide-mediated propagation between $j$-th and $k$-th emitters.
In this paper, we consider a standard simplification of the above model by choosing the regular distance between the emitters as an integer multiple of the waveguide's wavelength.
This allows us to effectively set the phases $\phi_{jk}$ and $\phi_{j}$ in the mediated interaction term and the jump operators to zero.
Additionally, in the fully chiral limit (we take $\gamma_L {=} 0$), the propagation phases can be gauged away within the Markovian descripion~\cite{carmichael2009statistical, pichler2015quantum}. Finite propagation delays beyond the Markov approximation give rise to genuine non-Markovian retardation effects~\cite{ramos2016non, Windt2025}; throughout this work, we assume the Markovian regime.

We will use the steady state of the master equation as the probe state to estimate the detuning strength of $\delta_j$.
We consider two types of detuning pattern: (i) uniform detuning with $\delta_j {=} h$ which enables frequency sensing; and (ii) Stark detuning with $\delta_j {=} jh$ that can be achieved by applying a gradient field to modify the emitter transition frequency and therefore can be used to enable frequency sensing.
In this paper, we set $\hbar {=} 1$ and $\gamma_R {+} \gamma_L {=} \Gamma$ as the unit of energy.

\section{Scaling of sensitivity}
\label{sec:scaling}

We now demonstrate the enhanced sensing capability by deriving the scaling of $\mathcal{F}^E$ with $N$, which establishes the lower bound for Fisher information.
For zero detuning, the steady state for even-$N$ is a pure dark state and can be written as product of dimer states~\cite{pichler2015quantum}, $\rho_{\rm ss}^{(0)} {=} \bigotimes_{n = 1,...,N/2} \ketbra{D_n}{D_n}$. 
Here the dimer state $\ket{D_n}$, consisting of emitters $2n{-}1$ and $2n$, can be represented on a 2-qubit dark state Bloch sphere by the spherical coordinates $(\eta_D, \theta_D, \phi_D) {=} (1, 2 \tan^{-1}{\frac{\sqrt{2} \Omega}{\gamma_R - \gamma_L}} , -\pi/2)$; see Appendix~\ref{app:Bloch} for details.

The two emitter dimer state for general detuning values, obtained by tracing out all the other emitters in the steady state, can be written down as a density operator with the dark state manifold and bright state manifold constituting the two diagonal blocks.
Evidently, for zero detuning, only the dark state manifold is populated.
If the detuning is turned on perturbatively, then within the linear response framework, one can track the correction within the dark state manifold.
While the population transfer to the bright state manifold will be later shown to be advantageous from the measurement perspective, we now focus on the dark state manifold to obtain the lower bound of the Fisher information.

With the two poles of the dark state Bloch sphere as $\ket{gg}$ and $\ket{\psi^-} {=} (\ket{eg} {-} \ket{ge})/\sqrt{2}$, we strategically build an effective average dark-dimer $X$-operator $X_D {=} \frac{1}{N/2} \sum_{n} \left(\ketbra{gg}{\Psi^-}_n {+} \ketbra{\Psi^-}{gg}_n \right)$.
To analytically compute the expectation value of $X_D$, one needs to restrict to the fully chiral case ($\gamma_L {=} 0$).
In this limit, owing to the fact that photons emitted from an emitter only affects another emitter to the right, the Liouvillian operator has a block-triangular matrix form (see Appendix~\ref{app:Heisenberg}).
We leverage this to solve the steady-state linear response recursively from the first dimer onwards. 
As detailed in Appendix~\ref{app:Green}, a perturbation on one dimer generates a local response that is relayed uniformly to all dimers to the right, yielding a simple lower-triangular response kernel from which $\langle X_D\rangle$ can be obtained analytically.
We find that for the $n$-th dimer, the spherical coordinates $\eta_D^{(n)}$ and $\theta_D^{(n)}$ remains unchanged to first-order of perturbation while $\phi_D^{(n)} {=} {-}\frac{\pi}{2} {+} \frac{(2\gamma_R^2+\Omega^2)}{\gamma_R(\gamma_R^2+2\Omega^2)} \sum_{k \le n} (\delta_{2k-1} {+} \delta_{2k})$.
This can be understood as an accumulation of azimuthal phase for each successive dimer (see Fig.~\ref{fig:schematic}(b)).
As shown in Appendix~\ref{app:scaling}, this leads to the sought-after scaling of $\mathcal{F}^E$ which, for the uniform detuning case ($\delta_j {=} h$), scales as $N^3$; and for the Stark detuning case ($\delta_j {=} jh$), this scales as $N^5$.
While these lower bounds of the scaling, derived at $h {=} 0$ and for the fully chiral case, establish the sensing advantage for very weak signals, we subsequently report numerical analysis of QFI and CFI to show the effectiveness in a broader range as well as robustness to finite chirality.

\section{QFI and CFI analysis}
\label{sec:FI}

\begin{figure}[t]
\centering
\includegraphics[width=0.49\textwidth]{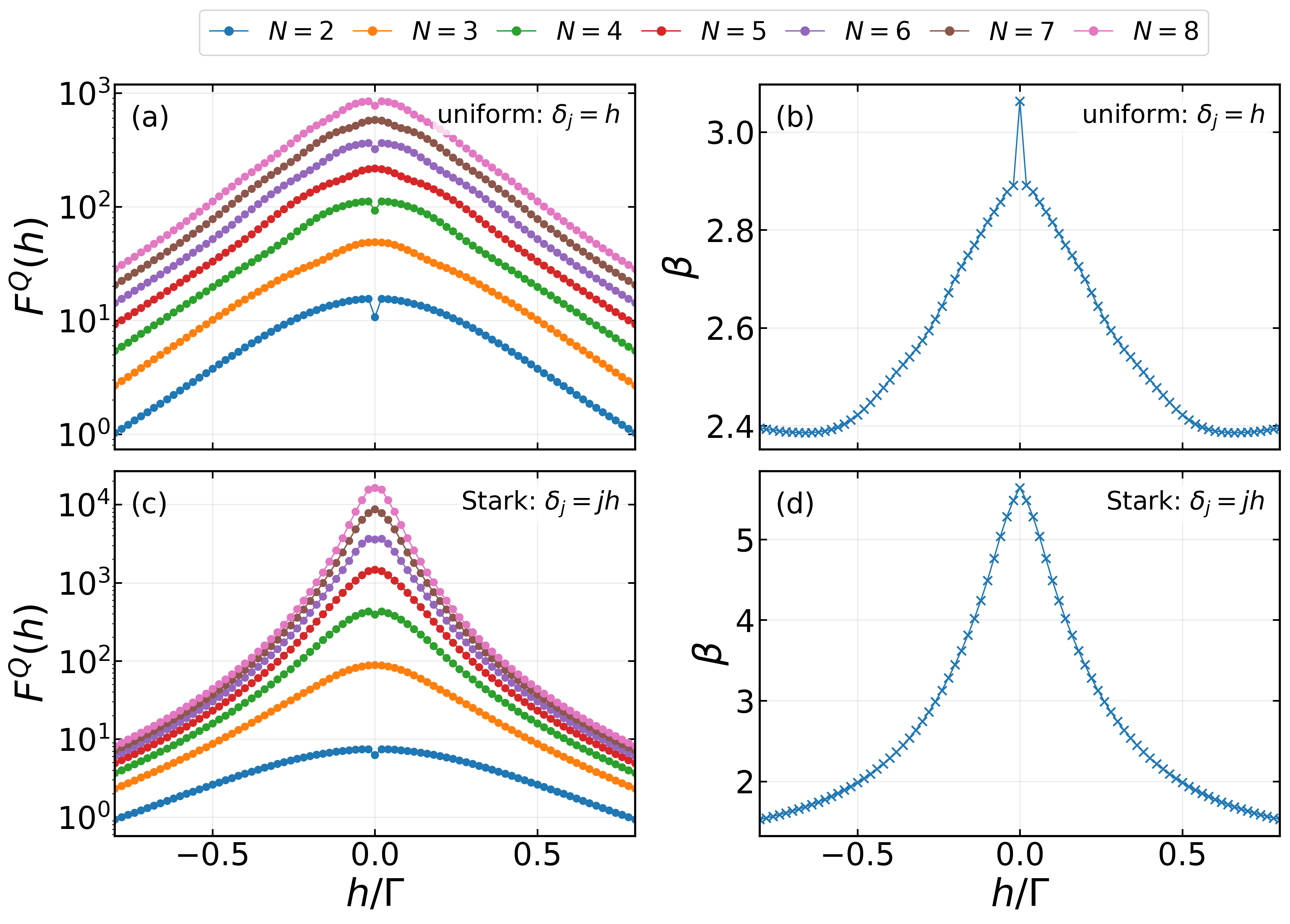}
\caption{\textbf{QFI analysis}. 
Top and bottom panel correspond to the uniform and Stark detuning, respectively.
QFI $F^Q$ as a function of the detuning strength $h$ for different system sizes with (a) uniform detuning and (c) Stark detuning.
Scaling exponent $\beta$ as a function of $h$ for (b) uniform detuning and (d) Stark detuning.
Here we fix $\Omega/\Gamma {=} 0.25$, $\gamma_R/\Gamma {=} 1$, and $\gamma_L/\Gamma {=} 0$.}
\label{fig:QFI}
\end{figure}

We now showcase the numerically calculated QFI values in a wider range of detuning strength $h$ for the fully chiral case.
As displayed in Fig.~\ref{fig:QFI}(a) for the uniform detuning case, the QFI profile shows a peak like structure centered at $h {=} 0$ which becomes more prominent as system size $N$ increases.
The scaling exponent $\beta$, obtained from fitting $F^Q {\sim} N^{\beta}$, reaches a value close to 3 at $h {=} 0$, similar to scaling behaviour of the lower bound.
As Fig.~\ref{fig:QFI}(b) shows, $\beta$ falls off gradually with increasing detuning strength but shows super-linear scaling in a wide range.
This shows the effectiveness of our probe state in terms of the range of applicability.
Similar behaviour is observed for the Stark detuning case as well, for which the QFI and scaling exponent are shown in Figs.~\ref{fig:QFI}(c) and (d), respectively.
While the numerical calculations are constrained by small system sizes, the relatively large values of the Fisher information are quite encouraging for the sensing capabilities.
For comparison, to obtain QFI value of $\mathcal{O}(10^3)$, it typically takes $\mathcal{O}(10^2)$ of two-level systems for a probe near quantum phase transition in the ground state of a transverse Ising chain~\cite{montenegro2021global} or in the steady state of a driven-dissipative system~\cite{montenegro2023quantum}. 

To compare the ultimate precision set by the QFI with that obtained in an experimentally friendly local measurement protocol, we compute the CFI for a staggered magnetization observable obtained from local Pauli $X$ operators, $M_x^{\rm stag} {=} \frac{1}{N} \sum_{j=1,...,N} (-1)^j X_j$ (see Appendix~\ref{app:CFI}).
As shown in Figs.~\ref{fig:CFI}(a) and (b) for the uniform and the Stark detuning scenario, respectively, the CFI closely follows the scaling laws obeyed by the QFI, although falling short of the optimal performance.

\begin{figure}[t]
\centering
\includegraphics[width=0.49\textwidth]{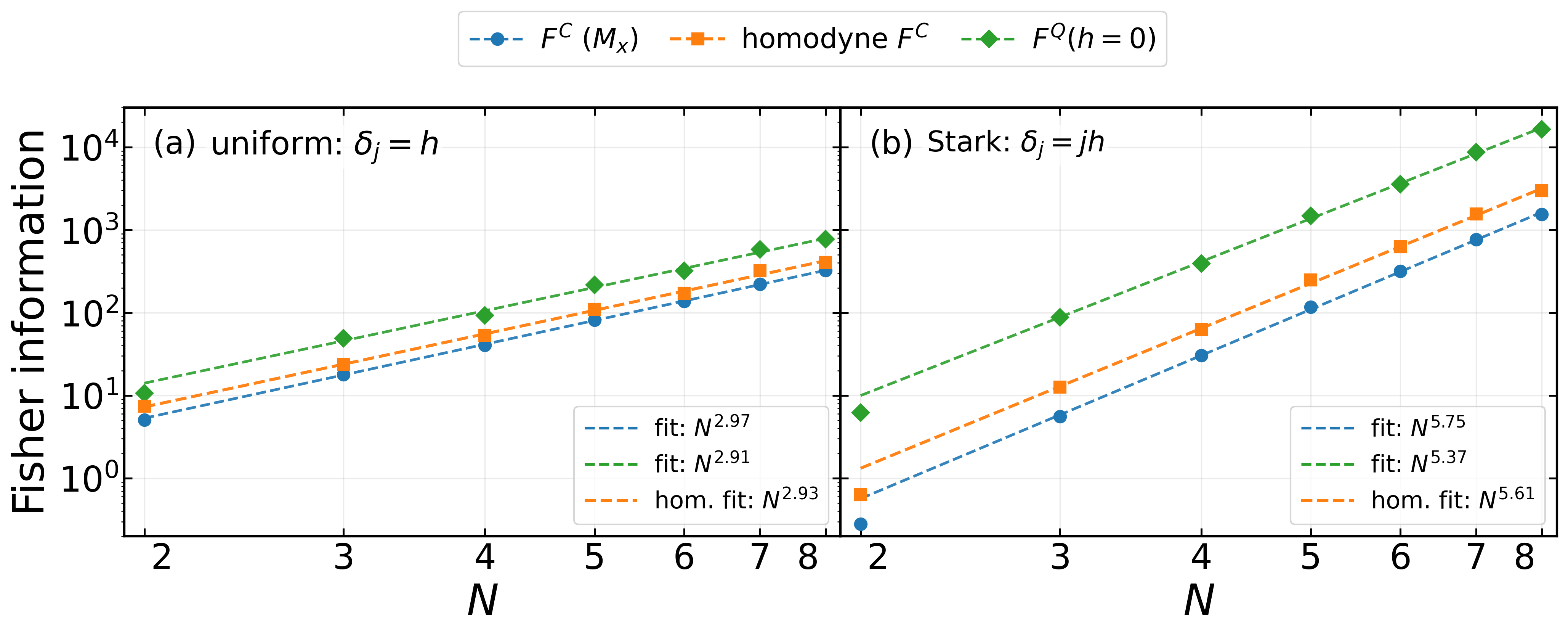}
\caption{\textbf{CFI analysis}. 
Comparison of the QFI with CFI calculated for the observable $M_x^{\rm stag}$ and CFI for the full-record homodyne scheme for the fully chiral case at $h {=} 0, \Omega/\Gamma {=} 0.25$ for the (a) uniform detuning and (b) Stark detuning.}
\label{fig:CFI}
\end{figure}

Another choice for calculating the CFI is given by homodyne detection measurements that are particularly relevant to our quantum optical system.
In this approach, one can directly monitor the output field of the chiral waveguide. 
We give an overview of the procedure here while more details can be found in Appendix~\ref{app:homodyne}.
Firstly, the steady state $\rho_{\rm ss}$ is subjected to Pauli $Z$ operators acting on alternative sites that changes the relative phase structure and increases population in the bright state manifold. 
Measurements are then carried out during a readout time $T_{\rm ro}$ at an interval $dt$.
The output field correspond to the bosonic operator $b_{\rm out}(t) = b_{\rm in}(t) + L_R(t)$ with $\braket{b_{\rm in}(t)}=0$ for vacuum input, and $L_R {=} \sqrt{\gamma_{R}} c_R$.
For balanced homodyne detection, a 50:50 beam splitter mixes $b_{\rm out}$ with a local oscillator with phase $\phi$ (quadrature operator $X_\phi {=} e^{-i\phi} L_R {+} e^{i\phi} L_R^\dagger$).
With total number of temporal bins $K {=} T_{\rm ro}/dt$, one can define the normalized bosonic output mode in bin $k$ by $b_k {=} \frac{1}{\sqrt{dt}} \int_{t_k}^{t_k+dt} b_{\rm out}(t)\,dt$ and the associated quadrature $Q_{k,\phi} {=} \frac{1}{\sqrt2} (e^{-i\phi} b_k {+} e^{i\phi} b_k^\dagger)$.
The measurements are carried out sequentially, in the POVM formed by the eigenvectors of $Q_{k,\phi}$ for the $k$-th instance, which gives a set of outcomes that define a trajectory.
The total probability of for a particular set of outcome is given by multiplying the individual conditional probabilities, which let us calculate the CFI.
We numerically simulate these process and average over $N_{\rm traj} {=} 300$ trajectories for the uniform and Stark detuning case and evaluate the CFI at $h{=}0$, while setting $\phi {=} 0$, $T_{\rm ro} {=} 5/\Gamma$, and $K {=} 50$.
As shown in Figs.~\ref{fig:CFI}(a) and (b), we again observe the scaling exponent of $F^C$ to be close to 3 and 5 respectively.

\section{Resource analysis}
\label{sec:time}

\begin{figure}[t]
\centering
\includegraphics[width=0.49\textwidth]{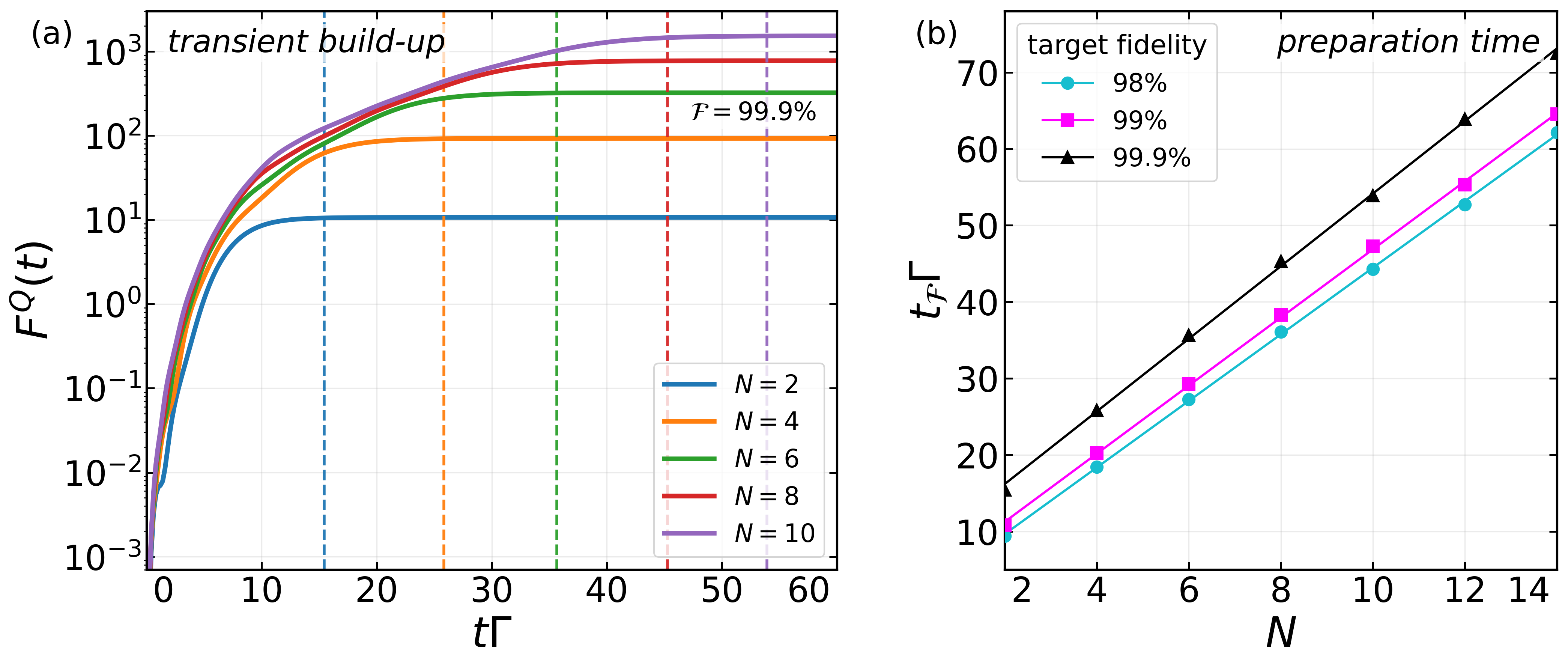}
\caption{\textbf{Probe preparation time}. 
(a) QFI $F^Q$ evaluated at $h{=}0$ of time for uniform detuning as a function of time as the system reaches the steady state.
Vertical dashed lines denote the time needed to reach $99.9\%$ of the steady state QFI value for each system size.
(b) Linear scaling of preparation time with system size for several choices of target fidelity with the steady state QFI value.}
\label{fig:time}
\end{figure}

While favourable scaling of steady state QFI exists for several other driven-dissipative systems near phase transitions, the time $T$ to prepare such critical states also generally scales with system size.
Therefore, it is more instructive to rather look at the scaling behaviour of $F^Q/T$ as a figure of merit~\cite{chaves2013noisy, brask2015improved, smirne2016ultimate, albarelli2018restoring, rossi2020noisy, sarkar2025first}.
In a closed quantum evolution governed by a Hamiltonian, such preparation time is associated with the energy gap~\cite{roland2002quantum, barankov2008optimal}.
For a master equation evolution, the Liouvillian gap can play a role~\cite{haga2021liouvillian, montenegro2023quantum}, but its definitive contribution is not guaranteed.
In fact, for our model, we find that Liouvillian gap is actually independent of the system size.
While numerical calculations confirm this for limited system sizes, we derive this analytically for the fully chiral case in Appendices~\ref{app:Heisenberg} and~\ref{app:time}.
The block triangular structure of the Liouvillian enables us to show that the constant gap is $-\gamma_R/2$ and that its generalised eigenspace for even $N$ consists of two parallel Jordan chain with length $N/2$.
Subsequently, we use this Jordan structure in Appendix~\ref{app:time} to show that the upper bound on the preparation time $T_\epsilon$ scales linearly with system size. 
Assuming that late-time dynamics are governed by the spectral-gap manifold, the longest-chain contribution is bound by $e^{-\gamma_Rt/2} (\gamma_R t)^{N/2-1}/(N/2-1)!$ and the physical transfer rates between successive Jordan layers are bounded by $\gamma_R$. 
Solving the corresponding relaxation-time bound and taking large $N$ asymptotics thus gives $T_\epsilon {\sim} \mathcal{O}(N/\gamma_R)$.
To confirm this numerically, we compute the QFI for the uniform detuning case with limited $N$ values as the system reaches steady state from an initial state with all emitters in the ground state.
We take the preparation time $T$ to be the time taken to reach $99.9\%$ of the steady state QFI value, as shown in Fig.~\ref{fig:time}(a).
This linear scaling of $T$ with $N$ is shown in Fig.~\ref{fig:time}(b), where we also show that slightly less fidelity with the steady state QFI value also has similar scaling.
This enables us to conclude that even with the preparation time taken into account, our system possesses enhanced sensing capability with at least $N^2$ scaling for the uniform detuning and $N^4$ scaling for the Stark detuning case.

\section{Effect of chirality}
\label{sec:chirality}

\begin{figure}[t]
\centering
\includegraphics[width=0.34\textwidth]{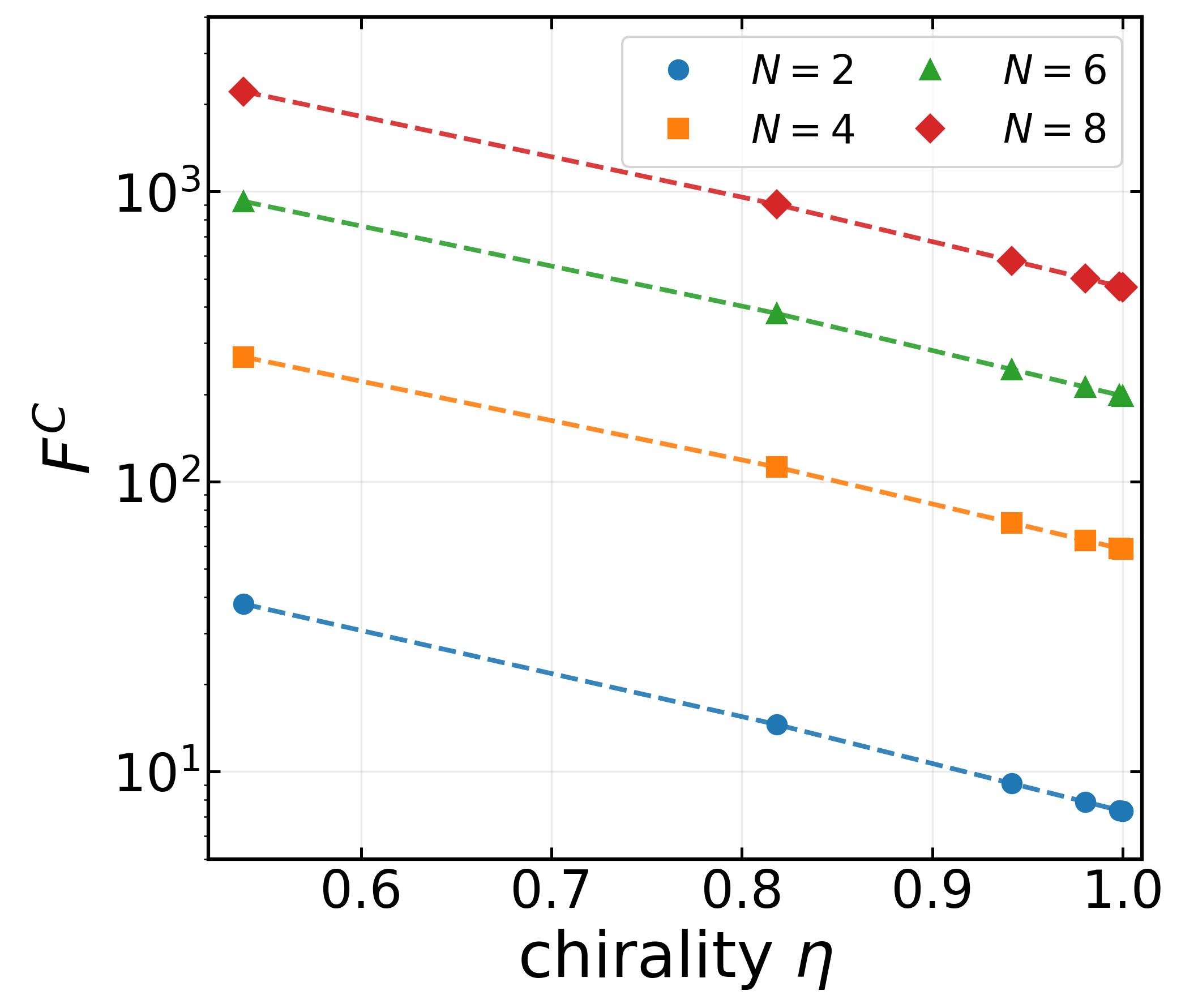}
\caption{\textbf{Effect of chirality}. Effect of chirality $\eta$ on the CFI evaluated for the observable $M_x^{\rm stag}$ at $h{=}0$ with uniform detuning for different system sizes.}
\label{fig:chiral}
\end{figure}

While we have mostly reported our findings for the fully chiral case due to analytical convenience, it is also important to study the effect of chirality $\eta {=} \frac{\gamma_R {-} \gamma_L}{\gamma_R {+} \gamma_L}$ on the sensing performance which can be below one in a physical realization.
Therefore we numerically evaluate $F^C$ for the observable $M_x^{\rm stag}$ at $h{=}0$ for the uniform detuning scenario and show the effect of $\eta$ in Fig.~\ref{fig:chiral}.
By analysing the scaling behaviour, we find that as $\eta$ decreases from 1 (fully chiral case), the scaling exponent slowly decreases but the prefactor increases.
This indicates that, for a given system size, one can obtain better precision by going away from the fully chiral limit.
However, we find that the range of the detuning strength over which the enhanced sensitivity is available, shrinks with decreasing chirality.
Therefore, the chirality provides us with a handle to trade between higher precision and higher range of enhanced sensing performance.

\section{Conclusion}
\label{sec:conclusion}

While dissipation is generally considered detrimental for quantum sensing, we explore it in this paper for generating probe states to achieve enhanced sensitivity.
By choosing the steady-state of an array of emitters connected to a chiral waveguide bath, we analytically show the enhanced sensitivity for estimating the detuning strength beyond quadratic scaling with system size while the probe preparation time only scales linearly.
This establishes an enhanced sensitivity even when the preparation time is considered.
We also provide experimentally implementable measurement methods to achieve sensitivity that are close to optimal.
Interestingly, the chirality of the waveguide acts as a handle for controlling the sensitivity and the range of sensing.
These results are encouraging for developing enhanced quantum sensors in the presence of a noisy environment.

\begin{acknowledgments}

AB acknowledges support from the National Natural Science Foundation of China (grants No.~W2541020, No.~12274059, No.~12574528, and No.~1251101297). 
LCK acknowledges support from the National Research Foundation and the Ministry of Education, Singapore.

\end{acknowledgments}

\appendix

\section{2-qubit steady-state at arbitrary detuning}
\label{app:Bloch}

For our purposes, it is convenient to describe a 2-qubit state in terms of Bloch spheres: a dark sector $D$ of $\mathcal{H}_D = \text{span}\{\ket{gg}, \ket{\Psi^-}\}$ and a bright sector $B$ of $\mathcal{H}_B = \text{span}\{ \ket{\Psi^+}, \ket{ee} \}$, where $\ket{\Psi^-} = (\ket{eg} - \ket{ge})/\sqrt{2}$ and $\ket{\Psi^+} = (\ket{eg} + \ket{ge})/\sqrt{2}$.
Decomposing an arbitrary 2-qubit mixed state thus results in:
\begin{equation}
    \rho=
    \begin{pmatrix}
        \rho_{DD} & C\\
        C^\dagger & \rho_{BB}
    \end{pmatrix}_{\mathcal{H}_D\oplus\mathcal{H}_B}
\end{equation}
where $C= P_D \rho P_B$ is the 2-by-2 correlation matrix accounting for dark-bright coherences, with $P$ being projectors onto respective spaces. Note that if $\rho$ is pure (i.e. Schmidt-rank 1), then $C$ can be represented by another Bloch-sphere with poles $\{\ket{eg}, \ket{ge}\}$.

Much of our analysis focuses only on the conditional state in the dark-sector. We define the coherence coordinate with respect to the Z-axis as $\alpha \in \mathbb{C}$, with the north-pole taken to be $\ket{gg}$ of the dark-Bloch sphere. From~\cite{pichler2015quantum}, if the two qubits experience opposite detunings $(\delta_1 = - \delta_2 = \chi)$ and uniform $\Omega, \gamma_{R(L)}$, then the two qubits settle into a pure dark-dimer state described by:
\begin{equation}
    \begin{split}
        \ket{D(\alpha)} &:= \frac{\alpha \ket{gg} + \sqrt{2} \ket{\Psi^-}}{\sqrt{|\alpha|^2 +2}}, \\
        \alpha &= \frac{2\chi}{\Omega} + \frac{i(\gamma_R - \gamma_L)}{\Omega}.
    \end{split}
\end{equation}

We can then decompose $\alpha$ into $r := \mathrm{Re}[\alpha] = 2 \chi/\Omega$ and $q := \mathrm{Im}[\alpha] = (\gamma_R - \gamma_L)/\Omega$.

For convenience, let us also define $\ket{T}$ and $\ket{T^*}$:
\begin{align}
    \ket{T(\alpha)} &:= \frac{\sqrt{2} \ket{gg} - \alpha^* \ket{\Psi^-}}{\sqrt{|\alpha|^2 +2}}, \\
    \ket{T^* (\alpha)} &:= \frac{\sqrt{2} \ket{ee} + \alpha^* \ket{\Psi^+}}{\sqrt{|\alpha|^2 +2}},
\end{align}
where $\ket{T}$ is antipodal to $\ket{D}$ on the dark-Bloch sphere, and $\ket{T^*}$ has the same angular coordinates as $\ket{T}$ but on the bright-Bloch sphere with its north-pole being $\ket{ee}$.

\subsection{Extension to arbitrary detuning}

The core mechanism behind the memory-sensor is that a common detuning shift leaks population into the bright-sector, creating an output field that is propagated downstream by the chiral waveguide. It is therefore of interest to characterize the two-qubit steady state under arbitrary detuning.

For brevity, we consider the chiral limit where $\gamma_L = 0$. Any $\delta_1, \delta_2 \in \mathbb{R}$ can be defined via two detuning components:
\begin{align}
    \chi &= \frac{\delta_1 - \delta_2}{2}, \\
    \Delta &= \frac{\delta_1 + \delta_2}{2},
\end{align}
where $\chi$ is the dimer-difference detuning and $\Delta$ is the dimer-common detuning.
We also generalize these quantities for the $j$-th dimer in a $N$-emitter system as
\begin{align}
    \chi_j &= \frac{\delta_{2j-1} - \delta_{2j}}{2}, \\
    \Delta_j &= \frac{\delta_{2j-1} + \delta_{2j}}{2} .
\end{align}

Making use of the shorthand: $u = \Omega/\sqrt{2}$ and $\zeta = -\chi + i\gamma_R/2$, we also predefine:
\begin{align}
    \mathcal{A} &= 2 \Delta ( \Delta^2 - \chi^2 + \gamma_R^2/4) + i\gamma_R ( \Delta^2 + \chi^2 + \gamma_R^2/4), \\
    \mathcal{Q} &= (\Delta^2 + u^2 -|\zeta|^2)^2 + \gamma_R^2 \Delta^2 + 4u^2|\zeta|^2, \\
    \mathcal{Z} &= (4\Delta^2 + \gamma_R^2) \mathcal{Q},
\end{align}
then the steady state is:
\begin{equation}
    \rho_{ss}= \frac{1}{\mathcal{Z}}
    \begin{pmatrix}
        m_{GG} & m_{GS} & m_{GT} & m_{GE}\\
        m_{GS}^* & m_{SS} & m_{ST} & m_{SE}\\
        m_{GT}^* & m_{ST}^* & m_{TT} & m_{TE}\\
        m_{GE}^* & m_{SE}^* & m_{TE}^* & m_{EE}
    \end{pmatrix},
\end{equation}
where the population entries are:
\begin{align}
    m_{SS} &= u^2 [(\Delta^2 + \gamma_R^2) u^2 + (4\Delta^2 + \gamma_R^2) |\zeta|^2],\\
    m_{TT} &= \Delta^2 u^2 (4\Delta^2 + \gamma_R^2 + u^2), \\
    m_{EE} &= \Delta^2 u^4, \\
    m_{GG} &= \mathcal{Z} - m_{SS} - m_{TT} - m_{EE},
\end{align}
and the coherence entries are:
\begin{align}
    m_{GS} &= u \zeta^* (2\Delta + i\gamma_R) [\mathcal{A} + u^2(-\Delta + i\gamma_R)] \\
    m_{GT} &= \Delta u (2\Delta + i \gamma_R) (\mathcal{A} + \Delta u^2) \\
    m_{GE} &= \Delta u^2 (2\Delta + i\gamma_R) (\Delta^2 + i\gamma_R \Delta - |\zeta|^2) \\
    m_{ST} &= \Delta u^2 \zeta (4\Delta^2 + \gamma_R^2) \\
    m_{SE} &= \Delta u^3 \zeta (2\Delta + i \gamma_R) \\
    m_{TE} &= \Delta^2 u^3 (2\Delta + i\gamma_R). 
\end{align}

Taking the conditional state within the dark-sector, 
\begin{equation}
    \rho_D = \frac{1}{m_{GG} + m_{SS}}
    \begin{pmatrix}
        m_{GG} & m_{GS} \\
        m_{GS}^* & m_{SS}
    \end{pmatrix},
\end{equation}
we can describe it with an $\alpha$-like stereographic coordinate together with $0 \leq \eta \leq 1$ describing the length of the conditional Bloch vector:
\begin{equation}
    \rho_D(\alpha_D, \eta_D) = \frac{1}{|\alpha_D|^2 +2}
    \begin{pmatrix}
        \frac{|\alpha_D|^2 + 2 + \eta_D (|\alpha_D|^2 -2)}{2} & \sqrt{2} \eta_D \alpha_D \\
        \sqrt{2} \eta_D \alpha_D^* & \frac{|\alpha_D|^2 + 2 - \eta_D (|\alpha_D|^2 -2)}{2}
    \end{pmatrix},
\end{equation}
which gives:
\begin{align}
    \eta_D &= \frac{D_D}{m_{GG} + m_{SS}}, \\
    \alpha_D &= \frac{2\sqrt{2} m_{GS}}{D_D - m_{GG} + m_{SS}},
\end{align}
where $D_D = \sqrt{(m_{GG} - m_{SS})^2+4|m_{GS}|^2}$.

Expanding around $\Delta = 0$, we find a perturbative expansion in $\Delta$:
\begin{equation}
    \alpha_D(\Delta) = \frac{2\chi + i \gamma_R}{\Omega} \left(1 - i \Delta \frac{ \Omega \mathcal{K}}{\gamma_R} \right) + \mathcal{O}(\Delta^2)
\end{equation}
where $\mathcal{K} = \frac{\Omega^2/2 + \gamma_R^2}{\Omega (\Omega^2/2 + \chi^2 + \gamma_R^2/4)}$. Note that to linear order, $\exp(-i\Delta\mathcal{K}/\gamma_R) \approx 1 - i\Delta\mathcal{K}/\gamma_R$; this can be interpreted as a rotation of $\alpha_D$.

Separating into real and imaginary components, we arrive at:
\begin{align}
    r_D &:= \text{Re}[\alpha_D] = \frac{2\chi}{\Omega} + \mathcal{K} \Delta + \mathcal{O}(\Delta^2), \\
    q_D &:= \text{Im}[\alpha_D] = \frac{\gamma_R}{\Omega} - \frac{2\chi \mathcal{K}}{\gamma_R} \Delta + \mathcal{O}(\Delta^2).
\end{align}

Thus, when all emitters are near-resonance with the waveguide $(\chi \ll \gamma_R/2)$ , a perturbation of $\Delta$ changes $\text{Re}[\alpha_{D}]$ at linear order but not $\text{Im}[\alpha_D]$. 

Importantly, we also note that a pure state remains pure to linear order of $\Delta$ since expanding around $\Delta = 0$ gives:
\begin{equation}
    \eta_D = 1 - \frac{\Omega^4}{2\gamma_R^2} \cdot \frac{(\chi^2 + \frac{\Omega^2 - \gamma_R^2}{4})^2 + \frac{3}{4}(\gamma_R^2 + \Omega^2/2)^2}{(\Omega^2/2 +\chi^2 + \gamma_R^2/4)^4} \Delta^2 + \mathcal{O}(\Delta^4),
\end{equation}
which then simplifies under $\gamma_R = 1$ and $\chi = 0$ to:
\begin{equation}
    \eta_D = 1 - \frac{8\Omega^4 (4\Omega^4 + 10 \Omega^2 + 13)}{(2\Omega^2 +1)^4} \Delta^2 + \mathcal{O}(\Delta^4).
\end{equation}

\subsection{Spherical coordinates}

Spherical coordinates make the geometry of the single-dimer response more explicit. Writing the conditional dark-sector state as $\rho_D^{cond} = \rho_D^{cond}(\eta_D, \theta_D, \phi_D)$, we use:
\begin{equation}
    \rho_D = \frac{1}{2}
    \begin{pmatrix}
        1 + \eta_D \cos \theta_D & \eta_D \sin \theta_D e^{-i\phi_D} \\
        \eta_D \sin \theta_D e^{i\phi_D} & 1 - \eta_D \cos \theta_D
    \end{pmatrix},
\end{equation}
where $\eta_D$ is the Bloch-vector radius, while $(\theta_D, \phi_D)$ specify angular direction. We arrive at:
\begin{align}
    \theta_D &= \arccos{\frac{m_{GG}-m_{SS}}{D_D}} = 2\arctan \frac{\sqrt2}{|\alpha_D|}, \\
    \phi_D &= - \arg m_{GS} = -\text{atan2}\,(\text{Im}[\alpha_D], \text{Re}[\alpha_D]).
\end{align}
Note that $\alpha_D = \sqrt{2}e^{-i\phi_D}\cot(\theta_D/2)$.

Expanding again about $\Delta = 0$ now yields,
\begin{align}
    \eta_D (\Delta) &= 1+ \mathcal{O}(\Delta^2),\\
    \theta_D(\Delta) &= 2 \arctan \frac{\sqrt{2} \Omega}{\sqrt{4\chi^2 + \gamma_R^2}} + \mathcal{O}(\Delta^2), \\
    \phi_D (\Delta) &= -\text{atan2}(\gamma_R, 2\chi) + \frac{\Omega\mathcal{K}}{\gamma_R} \Delta + \mathcal{O}(\Delta^2).
\end{align}
Thus the linear response to a dimer-common detuning $\Delta$ is tangent to the costant-radius, constant-polar angle circle of the dark-Bloch sphere: the entire first-order susceptibility is azimuthal.

At the resonant operating point $\chi = 0, \Delta = 0$, the azimuth is $\phi_0 = -\pi/2$. Further setting the fully right-chiral limit $\gamma_L=0$, we may define the dimensionless ratio, $q := \gamma_R/\Omega$, and evaluate the single-dimer azimuthal susceptibility at first-order:
\begin{equation}
    \beta_0 :=  \left. \frac{\partial \phi_D}{\partial \Delta} \right|_{\Delta = \chi = 0} = \frac{\Omega\mathcal K}{\gamma_R} = \frac{2(\Omega^2+2\gamma_R^2)}{\gamma_R(2\Omega^2+\gamma_R^2)} = \frac{2(2q^2+1)}{\gamma_R(q^2+2)}.
\end{equation}
Infinitesimally, the response can thus be simply summarised in local Bloch-ball coordinates as,
\begin{equation}
    (d\eta_D, d\theta_D, d\phi_D) = (0,0,\beta_0\,d\Delta) +\mathcal{O}(d\Delta^2).
\end{equation}
This geometric form will be the starting point for the many-dimer propagation argument in Appendix~\ref{app:Green}.

\section{Block-triangular Heisenberg-Liouvillian superoperator in the chiral limit: $\gamma_L = 0$}
\label{app:Heisenberg}

Recall that in the chiral limit ($\gamma_L = 0$), all waveguide-propagation phases in Eqs. \eqref{eq:master} and \eqref{eq:ham} can be gauged away~\cite{carmichael2009statistical}. It is then convenient to write the master equation in this form:
\begin{equation}
    \mathcal{L}[\rho] = -i ( H_{\rm eff}\rho - \rho H_{\rm eff}^\dagger ) + \mathcal{J}[\rho],
\end{equation}
where $H_{\rm eff}$ is the effective non-Hermitian Hamiltonian (or the generator at the no-jump limit), given by:
\begin{align}
    H_{\rm eff} &= H_{sys} - \frac{i\gamma_R}{2} \sum_j  \sigma_j^+ \sigma_j^- - i\gamma_R \sum_{j<k} \sigma_k^+ \sigma_j^-,
    \label{eq:Heff} \\
    H_{sys} &= \sum_j \left( -\delta_j \sigma_j^+ \sigma_j^- + \frac{\Omega_j}{2} \sigma_j^- + \frac{\Omega_j^*}{2} \sigma_j^+ \right),
\end{align}
and the recycling superoperator $\mathcal{J}$ collects all jump terms:
\begin{equation}
    \mathcal{J}[\rho] = \gamma_R \sum_{j,k} \sigma_j^- \rho \sigma_k^+ = J_R \rho J_R^\dagger,
    \quad J_R = \sqrt{\gamma_R} \sum_j \sigma_j^-.
\end{equation}

More pertinently, we note that in the chiral limit $\gamma_L = 0$, the master equation can also be rearranged into the superoperators:
\begin{equation}
    \mathcal{L}[\rho] = \sum_j \mathcal{L}_{j, (on)}[\rho] + \sum_{j<k}\mathcal{V}_{j \rightarrow k}^{R}[\rho]
\end{equation}
where the former term collects all lone-emitter contributions, 
\begin{equation}
    \mathcal{L}_{j, (on)}[\rho] = -i \left[ -\delta_j \sigma^+_j \sigma^-_j + \frac{\Omega_j}{2}\sigma_j^x , \rho \right] + \gamma_R \mathcal{D}[\sigma_j^-](\rho),
\end{equation}
while the latter contains a cascaded cross-emitter structure:
\begin{equation}
    \mathcal{V}_{j \rightarrow k}^{R}[\rho] = \gamma_R ([\sigma_j^- \rho, \sigma_k^+] + [\sigma_k^-, \rho \sigma_j^+]),
    \quad j<k.
\end{equation}

From this form, we would like to proceed to the Heisenberg picture, where we denote the Heisenberg-picture Liouvillian as $\mathcal{L}^\#$. By definition, $\mathcal{L}^\#$ and $\mathcal{L}$ are Hilbert-Schmidt adjoints that are related by the trace-dual condition~\cite{gardiner2004quantum}:
\begin{equation}
    \text{Tr}[O \mathcal{L}(\rho)] = \text{Tr}[(\mathcal{L}^\# O) \rho], \quad \forall O,\rho,
\end{equation}
for any operator $O$ and state $\rho$. 

Note that if, for computation, we choose to adopt a matrix-representation of $\mathcal{L}$ by vectorising operator-space with the column-stacking identity: $\text{vec}(AXB) = (B^T \otimes A)\, \text{vec}(X)$ for example; Then we have in the Schr\"odinger picture:
\begin{equation}
    \frac{d}{dt} \ket{\rho}\rangle = \mathbf{L}_S \ket{\rho}\rangle,
\end{equation}
where $\mathbf{L}_S$ is a $d^2 \times d^2$ matrix representing the superoperator $\mathcal{L}$, and the state-operator/density-matrix is mapped into a vector via column-stacking: $\rho \rightarrow \ket{\rho}\rangle \equiv \text{vec}(\rho)$. Hence the trace condition becomes:
\begin{equation}
    \langle \bra{O} \mathbf{L}_S \ket{\rho}\rangle = \langle \bra{O} \mathbf{L}_H^{\dagger} \ket{\rho}\rangle,
\end{equation}
where the matrices $\mathbf{L}_H$ and $\mathbf{L}_S$ are conjugate-transpose to each other.

Returning to the matter at hand, the Heisenberg-Liouvillian superoperator $\mathcal{L}^\#$ takes the simple form:
\begin{equation}
    \mathcal{L}^\# = \sum_j \mathcal{L}_{j, (on)}^\# + \sum_{j<k}\mathcal{V}_{j \rightarrow k}^{R\#}
\end{equation}
where the on-site term becomes:
\begin{equation}
    \begin{split}
        \mathcal{L}_{j, (on)}^\#  = &-i\, \left[-\delta_j \sigma_j^+ \sigma_j^- + \frac{\Omega_j}{2}(\sigma_j^+ + \sigma_j^-) , O \right] \\ &+ \gamma_R \left( \sigma_j^+ O \sigma_j^- - \frac{1}{2} \{\sigma_j^+ \sigma_j^- , O\} \right)
    \end{split}
\end{equation}
and the cross-emitter term now takes the form:
\begin{equation}
    \mathcal{V}_{j \rightarrow k}^{R\#}[O] = \gamma_R ([\sigma_k^+, O] \sigma_j^- + \sigma_j^+ [O, \sigma_k^-]),
    \quad j<k.
\end{equation}

Notice that $\mathcal{V}_{j \rightarrow k}^{R\#}[O]$ encourages a block-triangular matrix-representation for $\mathcal{L}^\#$: when an upstream site-$j$ emits an excitation, absorption by downstream site-$k$ generates an infinitesimal change in $O$ via the commutator, but this change is nonzero only if $O$ has (non-identity) support in site-$k$. The gist then is that $\mathcal{L}^\#$ can be represented as a matrix with block-triangular structure if we choose the basis in terms of observables whose most downstream support is site-$m$ and order them in ascending order: $1, 2, \cdots, N$.

The block-triangular structure of our Heisenberg-Liouvillian can be proven simply. Consider the generic observable denoted by the shorthand $O_{1, \cdots, m} = O_{1, \cdots, m} \otimes \mathbb{I}_{m+1, \cdots, N}$. It has non-trivial support only up to site-$m$.

Let $j<m<k$, so $O_{1, \cdots, m}$ commutes with operators acting on site-$k$. Thus,
\begin{equation}
    \mathcal{V}_{j \rightarrow k}^{R\#}[O_{1, \cdots, m}] = \gamma_R ([\sigma_k^+, O_{1, \cdots, m}] \sigma_j^- + \sigma_j^+ [O_{1, \cdots, m}, \sigma_k^-]) = 0.
\end{equation}
This means $\mathcal{L}^\#$ cannot expand the support of $O_{1, \cdots, m}$ downstream, i.e. under the action of $\mathcal{L}^\#$, $O_{1, \cdots, m} \rightarrow O_{1, \cdots, n}$ where $n \leq m$.

Hence, we can define the family of observables ordered by site-number $m$ as:
\begin{equation}
    \mathcal{A}_m := \mathcal{B}(\mathcal{H}_{1, \cdots, m}) \otimes \mathbb{I}_{m+1, \cdots, N},
\end{equation}
and we notice that they follow the filtration:
\begin{equation}
    \mathcal{A}_0 \subset \mathcal{A}_1 \subset \mathcal{A}_2 \subset \cdots \mathcal{A}_N,
    \quad \mathcal{A}_0 = z \cdot \mathbb{I}_{1, \cdots, N}, z \in \mathbb{C}.
\end{equation}
This means that,
\begin{equation}
    \mathcal{L}^\# \mathcal{A}_m \subseteq \mathcal{A}_m, \quad \forall m \in \{1, \cdots, N\}.
\end{equation}

Now if we construct a basis using the quotient spaces: $(Q_0, Q_1, Q_2, \cdots, Q_N)$, where $Q_m := \mathcal{A}_m / \mathcal{A}_{m-1}$ and $Q_0 = \mathcal{A}_0$, i.e. $Q_m$ is a set that consists of \textit{only} observables whose most downstream non-trivial support is on site-$m$, then we obtain the block-triangular matrix-representation:
\begin{equation}
    \mathcal{L}^\# = 
    \begin{pmatrix}
        B_0 & \bullet & \bullet & \cdots & \bullet \\
        0 & B_1 & \bullet & \cdots & \bullet \\
        0 & 0 & B_2 & \cdots & \bullet \\
        \vdots & \vdots & \vdots & \ddots & \vdots \\
        0 & 0 & 0 & \cdots & B_N
    \end{pmatrix},
\end{equation}
where the block-diagonal $B_m$ acts on quotient-space $Q_m$, while $\bullet$ denotes mixing from downstream sectors to upstream sectors. (This downstream-to-upstream relation seems like a contradiction, but recall that Heisenberg evolution is the adjoint of Schr\"odinger evolution; hence, an observable measured downstream at a later time is pulled back onto the upstream degrees of freedom that can influence it, whereas states and perturbations propagate physically from up- to downstream.)

One convenient basis choice for such a matrix is the Pauli-string basis ordered by rightmost non-trivial support: $\{\mathbb{I}_N\} \oplus \{X_1, Y_1, Z_1\} \oplus \{X_2, Y_2, Z_2, X_1X_2, X_1Y_2, X_1Z_2, Y_1X_2, \cdots\} \oplus \cdots$; identity operators have been suppressed for readability. We use this basis choice for the Heisenberg-Liouvillian moving forward.

Also, the above argument implies that a similar block-triangular matrix-representation can be found for the Schr\"odinger picture, but it is much less obvious: $\mathcal{V}_{j \rightarrow k}^{R}[\rho]$ describes how cross-emitter correlations develop when upstream site-$j$ emits and downstream site-$k$ absorbs an excitation, so ordering the basis via emitter state-levels generally gives a dense matrix.

\subsection{On eigenvalues and eigenoperators of $B_m$}

\begin{table}
    \centering
    \begin{tabular}{|c|c|c|}
        \hline
        $m$ & $\lambda^{\rm max}_m$ & $O^{\rm max}_m$ \\
        \hline
        1 & $-\gamma_R/2$ & $X_1$ \\
        \hline
        2 & $-\gamma_R/2$ & $Y_1 Z_2 - Z_1 Y_2$  \\
        \hline
        3 & $-\gamma_R/2$ & $Y_1 (Y_2 X_3 - X_2 Y_3) + Z_1 (Z_2 X_3 - X_2 Z_3)$ \\
        \hline
        4 & $-\gamma_R/2$ & $\begin{array}{c} Y_1 [X_2 (Z_3 X_4 - X_3 Z_4) + Y_2 (Z_3 Y_4 - Y_3 Z_4) ] \\+ Z_1[X_2 (X_3 Y_4 - Y_3 X_4) + Z_2 (Z_3 Y_4 - Y_3 Z_4)] \end{array}$ \\
        \hline
    \end{tabular}
    \caption{The maximum eigenvalue $\lambda^{\rm max}_m$ and the corresponding eigenvectors reshaped into eigen-operators $O^{\rm max}_m$ for each diagonal block $B_m$.}
    \label{tab:explicit-bm}
\end{table}

For illustration, we can explicitly show $B_1$:
\begin{equation}
    B_1 = 
    \begin{pmatrix}
        -\gamma_R/2 & \delta_1 & 0 \\
        -\delta_1 & -\gamma_R/2 & -\Omega_1 \\
        0 & \Omega_1 & -\gamma_R
    \end{pmatrix}.
\end{equation}
Let us set for all $j$, $\delta_j = 0$ and $\Omega_j = \Omega$ such that $0 < \Omega < \gamma_R/2$. 
Then the maximum non-zero eigenvalue of $B_1$ is $\lambda^{\rm max}_1 = -\gamma_R/2$, while its corresponding eigenoperator is $O^{\rm max}_1=X_1$.
Explicitly solving for several $B_m$, we note the corresponding $\lambda^{\rm max}_m$ and $O^{\rm max}_m$ in Table~\ref{tab:explicit-bm}.
The results suggest that the $\lambda^{\rm max}_m$ may be constant-in-$N$, with its corresponding eigenoperator following the implied rule:
\begin{equation}
    [\Sigma_1 \times (\Sigma_2 \times (\cdots \times \Sigma_m) \cdots)] := [x,y,z], \quad O^{\rm max}_m \propto  x.
\end{equation}
where we define the operator-valued vector $\Sigma_j = [X_j, Y_j, Z_j]$, with $\times$ denoting the vector cross-product.
$O^{\rm max}_m$ can be interpreted as chirally-dressed transverse coherences, but it should be noted these will not be eigenoperators of $\mathcal{L}^\#$; they have to be dressed by lower-support terms. 
On the other hand, it is known that the spectrum of a block-triangular matrix is the union of the spectra of its block-diagonals with repeat-entries allowed. 
This implies that the spectral gap of $\mathcal{L}$ is also constant-in-$N$, motivating Appendix~\ref{app:time}.

\section{Projected zero-frequency response and its causal propagation}
\label{app:Green}

This section derives the linear map from small dimer-common detunings $\Delta$ to the local dark-sector phase response used throughout the sensing analysis.

Appendix~\ref{app:Bloch} showed that, for an isolated dimer at resonance $\Delta = \chi = 0$, a perturbation in $\Delta$ produces a purely azimuthal displacement of the conditional dark-sector Bloch vector to first order: the Bloch radius $\eta$ and $\theta$ remains unchanged, while $\phi$ shifts. 

Although the system becomes nontrivial beyond two emitters because the waveguide mediates genuinely-correlated interactions beyond the dimer-local level, the many-body linear response can still be organised using the local dark-sector geometric picture introduced in Appendix~\ref{app:Bloch}. 

Firstly, strict chirality means that a perturbation on a given dimer-$k$ cannot influence dimers-$j$ when $j$ is upstream (i.e. $j<k$). Next, we derive the linear-response correction to the state and show it consists of a downstream single-dimer component and nearest-neighbour connected corrections with vanishing single-dimer marginals. Consequently, every dimer strictly downstream of a given source acquires the same reduced first-order correction. This combination of chirality and uniform downstream reduced response leads to the lower-triangular response kernel for the projected azimuthal displacement derived below.

We first prove the result for an even chain of $N=2M$ emitters grouped into $M$ dimers and for dimer-common detunings. The emitter-resolved and odd-$N$ extensions are given at the end. Throughout this section, we use the fully right-chiral limit $\gamma_L=0$ and keep the radiative rate $\gamma_R$ explicit. We write the dimensionless drive ratio as $q:=\gamma_R/\Omega$, while $\Delta_j$, $\chi_j$, and all other detunings retain their physical frequency units. Consequently, azimuthal susceptibilities such as $\beta_0$ have units of inverse frequency.

\subsection{Local Bloch-frame geometry and projected phase coordinate}

Recall from Appendix~\ref{app:Bloch} that for a dimer opposite-detunings $\chi \in \mathbb{R}$, the exact dark state is~\cite{pichler2015quantum}:
\begin{equation}
    \ket{D(\alpha)}= \frac{\alpha\ket{gg}+\sqrt{2}\ket{\Psi^-}}{\sqrt{|\alpha|^2+2}},
\end{equation}
with $ \alpha=\sqrt{2}e^{-i\phi}\cot(\theta/2)$. It is useful to embed this pure-state manifold in the conditional dark-sector Bloch sphere:
\begin{equation}
    \rho_D(\eta,\theta,\phi) = \frac{1}{2} \left[I+\eta\,\mathbf{n} (\theta,\phi) \cdot \boldsymbol{\sigma} \right],
\end{equation}
where $0 \leq \eta \leq 1$. We operate at resonance $(\Delta = \chi = 0)$ where the coordinates are,
\begin{equation}
    \eta_0 = 1,
    \quad \theta_0 = 2\arctan\frac{\sqrt2}{q},
    \quad \phi_0 = -\frac{\pi}{2},
\end{equation}
corresponding to $\alpha = iq$.

At this operating point, the local coordinate directions for the $j$-th dimer are represented by the tangent operators:
\begin{equation}
    R_{j,\mu} := \partial_{x_j^\mu} \rho_{D,j} (\eta_j, \theta_j, \phi_j) \Big\vert_{\Delta = \chi = 0},
    \quad x^\mu \in \{\eta, \theta, \phi\}.
\end{equation}

We note that these directions are mutually orthogonal in the Hilbert-Schmidt metric, with:
\begin{align}
    \text{Tr}[R_{\eta}^2] &= \frac{1}{2},
    \quad \text{Tr}[R_{\theta}^2] = \frac{\eta_0^2}{2},
    \quad \text{Tr}[R_{\phi}^2] = \frac{\eta_0^2 \sin^2 \theta_0}{2}, \\
    \text{Tr}[R_\mu R_\nu] &= 0 \quad (\mu \neq \nu).
\end{align}

This allows us to define the corresponding Hilbert-Schmidt dual covectors:
\begin{equation}
    W_j^\mu := \frac{R_{j,\mu}^\dagger}{\text{Tr}[R_{j,\mu}^\dagger R_{j,\mu}]},
    \quad \text{Tr}[W_j^\mu R_{j,\nu}] = \delta^\mu{}_{\nu}.
\end{equation}
The physical meaning is as follows. For an infinitesimal displacement within the conditional dark sector, we have,
\begin{equation}
    d\rho_{D,j} = R_{j,\eta} d\eta_j + R_{j,\theta} d\theta_j + R_{j,\phi} d\phi_j + \mathcal{O}(\norm{d\mathbf x_j}^2),
\end{equation}
so the trace-pairing directly returns the corresponding coordinate displacement:
\begin{equation}
    \text{Tr}[W_j^\mu \, d\rho_{D,j}] = dx_j^\mu + \mathcal{O}(\norm{d\mathbf x_j}^2).
\end{equation}
The geometric interpretation is thus clear: $R_{j,\mu}$ is a local tangent vector and $W_j^\mu$ is the dual covector that measures signed displacement along that coordinate.

Recall again from Appendix~\ref{app:Bloch} that at resonance, a perturbation in $\Delta$ selects only the azimuthal member of this frame to first-order:
\begin{equation}
    (d\eta_j, d\theta_j, d\phi_j) \approx (0, 0, \beta_0 \, d\Delta_j),
    \quad \beta_0 = \frac{2(2q^2 +1)}{\gamma_R (q^2 +2)},
\end{equation}
where $q = \gamma_R / \Omega$ is the dimensionless driving ratio. We therefore retain only $W_j^\phi$ in the response-propagation problem. 

For the many-body chain, we do not expect the first-order correction to the state to lie entirely within the local dark-manifold. Hence, if:
\begin{equation}
    \rho_{ss}(\vec{\Delta}) = \rho_0 + \rho^{(1)}(\vec{\Delta}) + \mathcal{O}(\Delta^2),
\end{equation}
where $\vec{\Delta} = \{\Delta_j\}_j$, then we can define the \textit{projected azimuthal displacement}:
\begin{equation}
    \varphi_j := \text{Tr}[W_j^\phi \rho^{(1)}(\vec{\Delta})].
\end{equation}
For an isolated dimer, this agrees with the literal azimuthal angle change to linear order; and in the full-chain, it remains a well-defined linear functional even when the response contains bright-sector or bright-dark correlation components.

Additionally, we note that at resonance, the azimuthal tangent and its dual can be express using a dimer-local `dark-Bloch effective-X' operator $\tau_j := \ketbra{gg}{\Psi^-}_j + \ketbra{\Psi^-}{gg}_j$:
\begin{equation}
    R_{j,\phi} = \frac{\sqrt2 q}{q^2 + 2} \tau_j,
    \quad W_j^\phi = \frac{q^2 +2}{2\sqrt2 q} \tau_j,
    \label{eqn: r-w-tau}
\end{equation}
which provides an intuitive starting point to define simple CFI-operators for sensing. This will be elaborated on in Appendix~\ref{app:scaling}.

Now with the above preamble settled, we can state the propagation relation and many-body response that underpins the sensing mechanism in our model. First note that at resonance (zero-detuning), the even-chain steady state factorises into a product of local adjacent dark-dimers:
\begin{equation}
    \rho_0 = \bigotimes_{j=1}^M \rho_{D,j}.
\end{equation}
Expressing the Schr\"odinger-Liouvillian as an expansion in $\Delta$:
\begin{equation}
    \mathcal{L}(\vec{\Delta}) = \mathcal{L}_0 + \sum_{k=1}^{M} \Delta_k \mathcal{L}_{1,k},
\end{equation}
we can define the generating-source produced by unit $\Delta$ on dimer-$k$ as:
\begin{equation}
     S_k := \mathcal{L}_{1,k}\rho_0 = i[N_k,\rho_0],
     \quad N_k = \sigma^+_{2k-1}\sigma^-_{2k-1} + \sigma^+_{2k}\sigma^-_{2k}.
\end{equation}

Let $\widetilde{\mathcal L}_0$ denote $\mathcal{L}_0$ on the complement of its stationary mode (equivalently, $\mathcal{L}_0$ with the steady-state component projected out). This allows us to define $G = - (\widetilde{\mathcal L}_0)^{-1}$ as the reduced resolvent for our system at zero-frequency. Making use of the steady-state relation $\mathcal{L}(\vec\Delta) \rho_{ss}(\vec\Delta) = 0$, we thus arrive at the first-order steady-state correction:
\begin{equation}
    \rho^{(1)} (\vec\Delta) = \sum_{k=1}^{M} \Delta_k G S_k.
\end{equation}

Thus, the projected Green's function is the azimuthal susceptibility matrix:
\begin{equation}
    G_{jk}^{(\phi)} := \text{Tr}[W_j^\phi G S_k],
    \quad \vec\varphi = G^{(\phi)} \vec\Delta.
\end{equation}
which provides $\vec{\varphi} = \{\varphi_j\}_j$ given some detuning profile $\vec\Delta = \{\Delta_k\}_k$ for a many-dimer chain.

The main result we will prove in this section is:
\begin{equation}
    G^{(\phi)}_{jk}=\beta_0
    \begin{cases}
        0,&j<k,\\
        1,&j=k,\\
        2,&j>k.
    \end{cases}
    \label{eq: projected-kernel-entry}
\end{equation}
The three entries have a direct interpretation: no upstream phase response, a local azimuthal response $\beta_0$, and a downstream response $2\beta_0$ after the disturbance has entered the chiral cascade.

\subsection{Propagation theorem}

It is convenient to prove Eq.~\eqref{eq: projected-kernel-entry} through a discrete spatial derivative. Let us define:
\begin{equation}
    D_j^\phi := W_j^\phi - W_{j-1}^\phi,
    \quad W_0^\phi = 0,
\end{equation}
and set $\Delta_0 = 0$. Then the following local statement is sufficient to prove:

\begin{theorem}
For an even chain of $M=N/2$ dark dimers at the resonant operating point,
\begin{equation}
    \text{Tr}[D_j^{\phi} G S_k] = \beta_0 \left( \delta_{jk} + \delta_{j-1,k} \right),
    \label{eq: local-propagation-theorem}
\end{equation}
where $\delta_{jk}$ is the Kronecker delta. Equivalently,
\begin{equation}
    \varphi_j - \varphi_{j-1} = \beta_0 (\Delta_j + \Delta_{j-1}).
    \label{eq: phi-recursion}
\end{equation}
\end{theorem}

In other words, a source farther upstream than dimer $j-1$ produces the same zero-frequency response on dimers $j-1$ and $j$, so it disappears from their difference. A downstream source cannot affect either observable because of chirality. Only the source on the dimer itself and the immediately upstream source remain. The proof of this thus relies on showing that all $k>j$ and $k<j-1$ terms cancel.

Additionally, letting $\varphi_0 = 0$ telescopes Eq.~\eqref{eq: phi-recursion} into:
\begin{equation}
    \varphi_j = \beta_0 \left( \Delta_j + 2\sum_{k<j} \Delta_k \right).
    \label{eq: phi-spatial-integrator}
\end{equation}
Hence, the chain acts as a zero-frequency spatial integrator: every detuning is recorded locally and is then carried to every downstream dimer. 

\subsubsection{Cancellation of $k>j$ terms: chirality}

To recap, Appendix~\ref{app:Heisenberg} shows that the Heisenberg-picture Liouvillian $\mathcal L^{\#}$ preserves the algebra of observables supported on an upstream prefix of the chain. Let $\mathcal{A}_{[1,n]}$ denote observables supported on sites $1,\cdots,n$ and identity downstream. Then we have,
\begin{equation}
    \mathcal{L}_N^{\#} \mathcal A_{[1,n]} \subseteq \mathcal A_{[1,n]}.
\end{equation}
Since $D_j^{\phi}$ is supported no farther downstream than dimer $j$, its reduced zero-frequency Heisenberg solution can also be chosen with support no farther than dimer $j$. Now we let $\Phi_j$ be the solution to:
\begin{equation}
    -\mathcal{L}_N^{\#} \Phi_j = D_j^\phi,
    \quad \text{Tr}[\Phi_j \rho_0] = 0.
\end{equation}
Next, we note that for some general operator $A$ and general state $\rho$, the Heisenberg-picture zero-frequency resolvent is given by: $G^\# = - (\widetilde{\mathcal L}_0^\#)^{-1}$ and is the Hilbert-Schmidt adjoint of $G$, thus resulting in the trace-relation after a bit of algebra:
\begin{equation}
    \text{Tr}[AG\rho] = \text{Tr}[(G^\# A) \rho].
    \label{eq:trace}
\end{equation}
Finally, we have:
\begin{equation}
    \text{Tr}(D_j^\phi G S_k) = \text{Tr}[(G^\# D_j^\phi) S_k] = \text{Tr}(\Phi_j S_k) = i \, \text{Tr}([\Phi_j, N_k] \, \rho_0).
\end{equation}
For $k > j$, $N_k$ is strictly downstream of $\Phi_j$, so the commutator vanishes. Hence,
\begin{equation}
    \text{Tr}[D_j^\phi G S_k] = 0,
    \quad k>j.
\end{equation}

\subsubsection{Cancellation of $k<j-1$ terms: uniform downstream reduced-response}

We next show why a source farther upstream than $j-1$ contributes equally to $W^{\phi}_{j-1}$ and $W^{\phi}_j$. To do so, we determine the structure of the full first-order steady-state correction generated by a single perturbative-source on dimer-$k$: $\rho^{(1,k)}$. 

The argument proceeds in three steps. First, we note that at resonance, the purity of the dark product-state allows the Liouvillian linear-response equation to be reduced to a linear equation involving the non-Hermitian effective Hamiltonian $H_{\rm eff}^{(M)}$, which is defined in Eq.~\eqref{eq:Heff}. Next, we solve for the effect of appending one unperturbed dark-dimer to the chain, and we obtain a two-dimer identity that preserves the form of the collective output correction while shifting its support to the newly appended dimer. Finally, we treat the prior solution as a base-case and extend it via induction to a length-$M$ chain. stream dimer, and nearest-neighbour connected contributions with vanishing one-dimer marginals. The equality of the downstream reduced corrections then gives the required cancellation in $D_j^\phi$.

To proceed, we first need to find the general form of $\rho_M^{(1)}$. Let us exploit the fact the resonant dark-state is pure. Letting $\ket{\mathcal{D}_M} = \ket{D}^{\otimes M}$ such that $\rho_0 = \ketbra{\mathcal{D}_M}{\mathcal{D}_M}$, we recall that the dark product obeys:
\begin{equation}
    H_{\rm eff}^{(M)} \ket{\mathcal{D}_M} = 0,
    \quad c\ket{\mathcal{D}_M} = \left(\sum_{m=1}^M C_m \right) \ket{\mathcal{D}_M},
\end{equation}
where $C_m = \sigma_{2m-1}^- + \sigma_{2m}^-$. Consequently, for any ket $\ket{\psi}$,
\begin{equation}
    \mathcal{L}_0 (\ketbra{\psi}{\mathcal{D}_M}) = -iH_{\rm eff}^{(M)} \ketbra{\psi}{\mathcal{D}_M}.
\end{equation}
The Hermitian-conjugate sector evolves analogously. Hence, operators of the form $\ketbra{\psi}{\mathcal{D}_M} + \ketbra{\mathcal{D}_M}{\psi}$ are mapped by $\mathcal{L}_0$ into operators of the same form.

Now we consider a unit common-detuning source on dimer-$k$, and we define:
\begin{equation}
    \ket{f_k} = (N_k - \bar{N}) \ket{\mathcal{D}_M},
\end{equation}
where $\bar{N} := \langle D \vert N_k \vert D \rangle = 2/(q^2 + 2)$. Note that $\langle \mathcal{D}_M \vert f_k \rangle = 0$.
From this, we receive:
\begin{equation}
    \mathcal{L}_{1,k} \rho_0 = i[N_k, \rho_0] = i(\ketbra{f_k}{\mathcal{D}_M} - \ketbra{\mathcal{D}_M}{f_k}).
\end{equation}
But recalling that the steady-state relation gives, $\mathcal{L}_{1,k} \rho_0 + \mathcal{L}_0 \rho_M^{(1,k)} = 0$, this implies that:
\begin{equation}
    \rho_M^{(1,k)} = \ketbra{u_M^{(k)}}{\mathcal{D}_M} + \ketbra{\mathcal{D}_M}{u_M^{(k)}},
    \label{eqn: pure-tangent}
\end{equation}
in which $\ket{u_M^{(k)}}$ is chosen in the gauge $\langle \mathcal{D}_M \vert u_M^{(k)} \rangle = 0$ to solve:
\begin{equation}
    H_{\rm eff}^{(M)} \ket{u_M^{(k)}} = \ket{f_k}.
\end{equation}

With the form of the first-order correction established, we now determine how it changes when a source-free dark dimer is appended downstream. Consider first a single dimer. Let $\ket{s}$ be the unique gauge-fixed solution of $H_{\rm eff}^{(1)} \ket{s} = \ket{f}$, and let us also define its state upon output into the waveguide: $\ket{p} = c\ket{s}$.

Now consider if we appended one source-free dark-dimer downstream. There ought to be some state $\ket{\xi}$ describing the exchange of excitation from the first dimer to the second:
\begin{equation}
    H_{\rm eff}^{(2)} \ket{\xi} = i\gamma_R (C\ket{s} \otimes C^\dagger \ket{D}) = i\gamma_R \ket{p} \otimes C^\dagger \ket{D}.
    \label{eqn: relay-xi}
\end{equation}
Explicitly solving for $\ket{\xi}$ gives rise to the following identity:
\begin{equation}
    \ket{p} \otimes \ket{D} + (C\otimes I + I \otimes C ) \ket{\xi} = \ket{D} \otimes \ket{p}.
    \label{eqn: relay-output}
\end{equation}

For completeness, we explicitly have:
\begin{align}
    \begin{split}
    \ket{s} &= \frac{2q}{\gamma_R (q^2+2)^{5/2}} \cdot [2(2q^2+1) \ket{gg} + iq(q^2 -1) \ket{ge} \\ &\qquad\qquad\qquad\qquad - 3iq(q^2 +1) \ket{eg} - (q^2 + 2) \ket{ee}]
    \end{split}
    \label{eqn: s-explicit}
    \\
    \ket{p} &= - \frac{2q}{\gamma_R (q^2+2)^{3/2}} \cdot (2iq \ket{gg} + \ket{eg} + \ket{ge}).
    \label{eqn: p-explicit}
\end{align}

It is also useful to make the decompose $\ket{\xi}$ into a factorised component and a connected component:
\begin{equation}
    \ket{\xi} = \ket{D} \otimes \ket{v} + \ket{w},
    \label{eqn: xi-comp}
\end{equation}
under the condition that: 
\begin{equation}
    (\bra{D} \otimes I) \ket{w} = (I \otimes \bra{d}) \ket{w} = 0.
    \label{eqn: w-condition}
\end{equation}
Here, using the basis $\{ \ket{gg}, \ket{ge}, \ket{eg}, \ket{ee} \}$, the one-dimer downstream component $\ket{v}$ is:
\begin{equation}
    \begin{split}
    \ket{v} = \frac{1}{\gamma_R} \cdot \Bigg\{\frac{8q(2q^2+1)}{(q^2+2)^{5/2}}, -\frac{4iq^2(2q^2+1)}{(q^2+2)^{5/2}} ,\\ \frac{4iq^2(2q^2+1)}{(q^2+2)^{5/2}}, -\frac{4q}{(q^2+2)^{3/2}} \Bigg\}.
    \end{split}
    \label{eqn: relay-v}
\end{equation}
While the vector $\ket{w}$ is a connected nearest-dimer term required for the two-dimer solution, its contribution vanishes when taking the marginals of either single-dimer and therefore does not contribute to either's local reduced response. Higher-order correlations do not appear presumably since we operate to first-order in the perturbation.

Thus, Eqs.~\eqref{eqn: relay-xi} and~\eqref{eqn: relay-output} give the two-dimer extension identities needed for the induction to a chain of arbitrary length. We claim that for every $M \geq 1$, there exists a response vector $\ket{u_M}$ such that the following two relations hold:
\begin{align}
    H_{\rm eff}^{(M)} \ket{u_M} &= \ket{f} \otimes \ket{D}^{\otimes (M-1)}, 
    \label{eqn: relay-induc-H}
    \\
    c_M \ket{u_M} &= \ket{D}^{\otimes (M-1)} \otimes \ket{p}.
    \label{eqn: relay-induc-C}
\end{align}
Note that we have placed the source on the first dimer; a source at general $k$ follows trivially since dark-dimers can be tensored upstream to no significance due to chirality.

The base case of this process is $\ket{u_1} = \ket{s}$ and has been established above. For the inductive step, assume the two relations hold for some $M$. To extend to $M+1$, strict chirality gives us:
\begin{equation}
    H_{\rm eff}^{(M+1)} =  H_{\rm eff}^{(M)} \otimes I + I \otimes H_{\rm eff}^{(1)} - i\gamma_R C_{M+1}^\dagger c_M.
    \label{eqn: relay-Heff-append}
\end{equation}
We can also use the recursive ansatz:
\begin{equation}
    \ket{u_{M+1}} = \ket{u_M} \otimes \ket{D} + \ket{D}^{\otimes (M-1)} \otimes \ket{\xi}.
    \label{eqn: relay-u-append}
\end{equation}

Using Eq.~\eqref{eqn: relay-induc-C}, the last term in Eq.~\eqref{eqn: relay-Heff-append} acting on $\ket{u_M}\otimes\ket{D}$ produces $-i\gamma_R \ket{D}^{\otimes(M-1)} \otimes \ket{p} \otimes c^\dagger \ket{D}$, which is precisely canceled by Eq.~\eqref{eqn: relay-xi}. This proves Eq.~\eqref{eqn: relay-induc-H} at $M+1$.

Likewise, we have:
\begin{equation}
    c_{M+1} \ket{u_{M+1}} = \ket{D}^{\otimes(M-1)} \otimes [\ket{p} \otimes \ket{D} +(c \otimes I + I \otimes c) \ket{\xi}],
\end{equation}
which combined with Eq.~\eqref{eqn: relay-output} proves Eq.~\eqref{eqn: relay-induc-C} at $M+1$. Thus, the same collective-output correction is reproduced on the terminal dimer after each extension of the chain.

More importantly, this validates the ansatz used in Eq.~\eqref{eqn: relay-u-append} and makes the structure of $\ket{u_M}$ explicit (for single source on dimer-$1$):
\begin{equation}
    \begin{split}
        \ket{u_M} = \ket{s}_1 \ket{D}^{\otimes (M-1)} &+ \sum_{m=2}^{M} \ket{D}_1 \cdots \ket{v}_m \cdots \ket{D}_M \\
        &+ \sum_{m=1}^{M-1} \ket{D}_1 \cdots \ket{w}_{m,m+1} \cdots \ket{D}_M.
    \end{split}
\end{equation}
Thus, to first-order, $\ket{u_M}$ contains only a local tangent $\ket{s}$ at the source-dimer, the same one-dimer downstream tangent $\ket{v}$ on every later dimer, and identical nearest-dimer connected tangents $\ket{w}$ on the intervening bonds; no longer-ranged connected cluster is required. 

Somewhat counterintuitively, the perturbation is therefore transmitted over arbitrarily many dimers through a sequence of local two-dimer adjustments, rather than through the formation of a long-range correlated cluster as we might expect from a typical equilibrium system.

It thus follows from this and Eqs.~\eqref{eqn: pure-tangent} and~\eqref{eqn: xi-comp} that every dimer strictly downstream of a source on some arbitrary dimer-$k$ will have the same reduced first-order correction, and will not incur the connected-term unless $k = j-1$. Thus,
\begin{equation}
    \rho_{m}^{(1,k)} = \ketbra{v}{D} + \ketbra{D}{v}, 
    \quad k < j-1.
    \label{eqn: equal-downstream-reduced}
\end{equation}
Therefore, a source $k<j-1$ contributes equally to the two neighboring projected-phase readouts:
\begin{equation}
    \text{Tr}[W_j^{\phi} G S_k] = \text{Tr}[W_{j-1}^{\phi} G S_k], 
\end{equation}
and hence,
\begin{equation}
    \text{Tr}[D_j^{\phi} G S_k] = 0,
    \quad k < j-1.
\end{equation}

\subsubsection{Explicit solving for $k = j$ and $k = j-1$ terms}

The final two cases can be solved explicitly. For $k=j$, the system is effectively one isolated dimer. Linking back to Eqs.~\eqref{eqn: equal-downstream-reduced} and~\eqref{eqn: relay-v}, we note that the trace-zero solution for $-\mathcal{L}_D \rho_1^{(1,1)} = S$ is:
\begin{equation}
    \begin{split}
    \rho_1^{(1,1)} &= \ketbra{s}{D} + \ketbra{D}{s} \\
    &= \frac{q}{\gamma_R (q^2 + 2)^2}
    \begin{pmatrix}
        0 & -2(3q^2 + 1) & 2(q^2 + 1) & -2iq\\
        -2(3q^2 + 1) & 0 & -4iq & 2 \\
        2(q^2 + 1) & 4iq & 0 & -2 \\
        2iq & 2 & -2 & 0
    \end{pmatrix},
    \label{app:eqn:G}
    \end{split}
\end{equation}
using the basis $\{ \ket{gg}, \ket{ge}, \ket{eg}, \ket{ee} \}$.

Contracting with $W^\phi$ gives,
\begin{equation}
    \text{Tr}[W\nu] = \beta_0 = \frac{2(2q^2 + 1)}{\gamma_R(q^2 + 2)}.
\end{equation}
This is the same local susceptibility obtained by expanding the exact single-dimer solution of Appendix~\ref{app:Bloch} at $\chi = 0$.

For $k=j-1$, we solve the two-dimer relay problem where the source is on dimer-$(j-1)$. Relabelling dimer-$(j-1)$ as -$1$ and dimer-$j$ as -$2$, we arrive at:
\begin{equation}
    \begin{split}
            \rho_2^{(1,1)} = \, &(\ketbra{s}{D} + \ketbra{D}{s})_1 \otimes \rho_{D,2}\\ 
            &+ \rho_{D,1} \otimes (\ketbra{v}{D} + \ketbra{D}{v})_2 \\ &+ \ketbra{w_{12}}{D_1 D_2} \otimes \ketbra{D_1 D_2}{w_{12}}.
    \end{split}
\end{equation}
By exact solving, the projected response is:
\begin{equation}
    \text{Tr}[W_m^\phi \rho_2^{(1,n)}] = \beta_0 \cdot
    \begin{cases}
        0, & m<n, \\
        1, & m=n, \\
        2, & m>n.
    \end{cases}
\end{equation}

Thus, a source on dimer $j-1$ contributes $2\beta_0$ to dimer-$j$ and $\beta_0$ to dimer-$(j-1)$, so their difference is again $\beta_0$. Together with the two vanishing cases above for $k>j$ and $k<j-1$ terms, this proves Eq.~\eqref{eq: local-propagation-theorem}.

\subsubsection{Generic projected response form}

Since we know the generic form of the first-order response $\rho_M^{(1,k)}$ where the source is on dimer-$k$, we can define the projected response more generically for any dimer-local operator $O$ as:
\begin{equation}
    G_{jk}^{(O)} := \text{Tr}[O_j \rho_{M}^{(1,k)}] = \text{Tr}[O_j GS_{k}].
\end{equation}
Recall from Eq.~\eqref{eqn: w-condition} that terms with $\ket{w}$ have zero dimer-local marginal. Therefore, we have:
\begin{equation}
    G_{jk}^{(O)} =
    \begin{cases}
        0, & j < k, \\
        2\, \text{Re} \langle D \vert O \vert s \rangle, & j=k, \\
        2\, \text{Re} \langle D \vert O \vert v \rangle, & j>k.
    \end{cases}
    \label{eqn: generic-proj-resp}
\end{equation}

As an aside, suppose we have a two-dimer operator $\Xi$ with support on adjacent dimers-$j$ and -$(j+1)$. Then we have:
\begin{equation}
    G_{jk}^{(\Xi)} =
    \begin{cases}
        0, & k > j+1, \\
        2\, \text{Re} \langle DD \vert \Xi \vert Ds \rangle, & k=j+1, \\
        2\, \text{Re} \langle DD \vert \Xi (\ket{sD} + \ket{Dv} + \ket{w}), & k=j, \\
        2\, \text{Re} \langle DD \vert \Xi (\ket{Dv} + \ket{vD} + \ket{w}), & k<j,
    \end{cases}
\end{equation}
though sensing advantages in using a two-dimer operator are not expected.

\subsection{Matrix form and projected Jordan structure}

Note that Eq.~\eqref{eq: phi-spatial-integrator} can also be written as:
\begin{equation}
    \vec\varphi = \beta_0 K_M \vec\Delta,
\end{equation}
where:
\begin{equation}
    [K_M]_{jk}=
    \begin{cases}
        0, & k>j,\\
        1, & k=j,\\
        2, & k<j.
    \end{cases}
\end{equation}

Now let $S_M$ be the downstream shift operator such that $S_M e_j = e_{j+1}$ and $S_M e_M = 0$, where $e_j$ denotes the $j$-th dimer in position basis, $j=1$ being most upstream. Since $S_M^M =0$,
\begin{equation}
    K_M = (I+S_M)(I-S_M)^{-1} = I + 2\sum_{p=1}^{M-1} S_M^p.
\end{equation}
The term $S_M^p$ represents propagation by exactly $p$ dimers. Every propagation depth carries the same zero-frequency weight, which is a direct consequence of the transparent-relay property established earlier.

Additionally, we note that $(K_M - I)^M = 0$ while $(K_M - I)^{M-1} \neq 0$, so $K_M$ has minimal polynomial $(\lambda - 1)^M$ and is similar to a Jordan block of length-$M$ and eigenvalue 1. Consequently, the projected susceptibility $G^{(\phi)}=\beta_0K_M$ has a single projected Jordan chain with eigenvalue $\beta_0$. Of note, this statement concerns the  projected zero-frequency response map rather than the full Liouvillian; while the Jordan structure in the full Liouvillian has implications on the preparation time to steady-state as discussed in Appendix~\ref{app:time}, the Jordan structure in the response map clarifies the spatial accumulation underpinning the sensing mechanism and is used to derive the corresponding CFI lower bound in Appendix~\ref{app:scaling}.

\subsection{Emitter-resolved detuning}

Thus far, we have focused on dimer-resolved detuning (i.e. where detuning levels are the same between both emitters of a dimer), but it is natural to consider emitter-resolved detuning instead. 
With the dimer-average detuning $\Delta_j$ and the dimer-opposite detuning $\chi_j$ for the $j$-th dimer, we note that when $\chi_j = 0$, $\Delta_j$ is equivalent to dimer-common detuning like before.
The average component $\Delta_j$ propagates through the chiral kernel derived above, whereas the dimer-opposite component directly shifts the local dark-state azimuth. 
Since $r = q\varphi + \mathcal{O}(\varphi^2)$ at the operating point, while the exact isolated-dimer dark-state coordinate gives $r = 2\chi / \Omega = 2q \chi / \gamma_R$, the antisymmetric detuning produces $\varphi = 2\chi / \gamma_R$ to linear order. 
Hence,
\begin{equation}
    \vec{\varphi} = \beta_0 K_M \vec\Delta + \frac{2}{\gamma_R} \vec{\chi}.
\end{equation}

\subsection{Odd-N chains}

For $N=2M+1$, the zero-detuning steady state is simply:
\begin{equation}
    \rho_0^{(2M+1)} = \rho_D^{\otimes M} \otimes \rho_\ell,
\end{equation}
with terminal-emitter state,
\begin{equation}
    \rho_\ell = \frac{1}{q^2+1}
    \begin{pmatrix}
        1&-iq\\
        iq&q^2+1
    \end{pmatrix}.
\end{equation}
The terminal emitter is downstream of the unchanged even-$N$ dimer block.

If $x_\ell$ denotes the chosen linear-response coordinate of that terminal emitter, the previously obtained one-emitter/two-block solve gives:
\begin{equation}
    \begin{pmatrix}
        \vec\varphi \\
        x_\ell
    \end{pmatrix}
    =
    \begin{pmatrix}
        \beta_0 K_m & 0 \\
        g_{\ell r} \mathbf{1}_M^T & g_{\ell \ell}
    \end{pmatrix}
    \begin{pmatrix}
        \vec\Delta \\
        \delta_\ell
    \end{pmatrix},
\end{equation}
where we have:
\begin{equation}
    g_{\ell r} = \frac{16(q^2+1)}{\gamma_R(q^2 + 2)^2},
    \quad g_{\ell \ell} = \frac{4q}{\gamma_R (q^2 + 2)}.
\end{equation}

Accounting for emitter-resolved detuning changes this to:
\begin{equation}
    \begin{pmatrix}
        \vec\varphi \\
        x_\ell
    \end{pmatrix}
    =
    \begin{pmatrix}
        \beta_0 K_m & \frac{2}{\gamma_R}I_M & 0 \\
        g_{\ell r} \mathbf{1}_M^T & 0 & g_{\ell \ell}
    \end{pmatrix}
    \begin{pmatrix}
        \vec\Delta \\
        \vec\chi \\
        \delta_\ell
    \end{pmatrix}.
\end{equation}

\subsection{Validity and robustness of sensing regime}

For the trivial case of a single dimer, the sensing regime is valid for $|\beta_0 \Delta | \ll 1$ which is parametrically similar to $|\Delta| \ll \gamma_R$. The many-dimer case is not as straightforward. Even if every dimer-$j$ satisfies $|\Delta_j| \ll \gamma_R$, when $N$ becomes very large, the accumulated azimuthal phase $\varphi_j$ can eventually become large too.

Thus, for us to remain within the linear-response locally-pure regime, the actual condition is:
\begin{equation}
    \max_j{ \Bigg\vert \beta_0 \left( \Delta_j + 2\sum_{k<j} \Delta_k \right) \Bigg\vert} \ll 1.
\end{equation}

For a uniform pattern ($\Delta_j = \Delta$), the largest phase occurs on the final dimer: $\varphi_M = \beta_0 (2M-1) \Delta$, which alludes to $\beta_0 (2M-1) |\Delta| \ll 1$, and hence $N |\Delta| \ll \gamma_R$.

\section{Analytic lower bound for Fisher information scaling}
\label{app:scaling}

Building off the propagation theorem, we analytically derive a lower bound for our model-specific QFI at even-N and show it scales as $N^3$ for uniform detuning. Specifically, we do so by defining a readout observable for classical Fisher information and use the error-propagation formula to derive its N-scaling. Numerically we also find that the CFI almost saturates the scaling of QFI.

Consider detuning perturbations around our operating point at $\delta_j = 0$. Appendix~\ref{app:Green} identifies the projected azimuthal displacement $\varphi_j$ as the natural local sensing coordinate, so an intuitive scalar response for readout is the average accumulated phase:
\begin{equation}
    \bar\varphi := \frac{1}{M} \sum_{j=1}^{M} \varphi_j,
\end{equation}
where $M = N/2$ is the number of dimers.

To read out this phase, we first define the dark-dimer effective-X operator:
\begin{equation}
    \tau_j = \ketbra{gg}{\Psi^-}_j + \ketbra{\Psi^-}{gg}_j,
\end{equation}
and recall from Eq.~\eqref{eqn: r-w-tau} that:
\begin{equation}
    W_j^\phi = \frac{q^2 + 2}{2\sqrt{2} q} \tau_j.
\end{equation}

Hence, by the definition $\varphi_j = \text{Tr}[W_j^\phi \rho^{(1)}]$, any first-order response satisfies:
\begin{equation}
    \text{Tr}[\tau_j \rho^{(1)}] = \frac{2\sqrt2 q}{q^2+2} \varphi_j,
\end{equation}
or equivalently,
\begin{equation}
    \partial_h \langle \tau_j \rangle \vert_{h=0} = \frac{2\sqrt2 q}{q^2+2} \partial_h \varphi_j \vert_{h=0}.
    \label{eqn: phi-to-tau}
\end{equation}
Hence, $\tau_j$ is a transverse Bloch-sphere quadrature that converts the projected azimuthal displacement into a directly measurable signal.

Let us then define the associated average over the chain:
\begin{equation}
    X_D = \frac{1}{M} \sum_{j=1}^{M} \tau_j.
\end{equation}
Its expectation value reads out $\bar\varphi$ with a prefactor  $2\sqrt2q/(q^2+2)$.

By the error-propagation formula, for any observable $O$,
\begin{equation}
    F_{Q,\theta} \geq F_{C,\theta}[M] \geq \frac{\lvert \partial_\theta \langle O \rangle \rvert^2}{\text{Var}(O)}.
\end{equation}

At our operating point $\delta_j = 0$, the exact steady-state is $\rho_0 = \rho_D^{\otimes M}$. For a single dimer, we get $\langle \tau \rangle_0 = 0$. Moreover, $\tau^2 = \ketbra{gg}{gg} + \ketbra{\Psi^-}{\Psi^-}$ is the projector to the dark-sector, so $\langle \tau^2 \rangle_0 = 1$ and $\text{Var}_{\rho_D}(\tau) = 1$. Since the steady-state factorises, we thus have,
\begin{equation}
    \text{Var}_{\rho_0}(X_D) = \frac{1}{M}.
\end{equation}

Next, the numerator depends on the detuning-perturbation profile we choose. For a dimer-resolved uniform detuning profile of $\Delta_j = h$ for $h\in \mathbb{R}$, the propagation theorem gives:
\begin{equation}
    \partial_h \varphi_j \big\vert_{h=0} = \beta_0 (2j-1).
\end{equation}
Using Eq.~\eqref{eqn: phi-to-tau},
\begin{equation}
    \begin{split}
        \partial_h \langle X_D \rangle \vert_{h=0} &= \frac{2\sqrt{2}q}{M (q^2 + 2)} \sum_{j=1}^{M} \beta_0 (2j-1) \\
        &= \frac{2\sqrt{2}q \beta_0}{q^2 + 2} M,
    \end{split}
\end{equation}
so the error-propagation quantity evaluates to:
\begin{equation}
    \frac{\lvert \partial_\theta \langle X_D \rangle \rvert^2}{\text{Var}(X_D)} = \frac{8q^2 \beta_0^2}{(q^2 + 2)^2} M^3.
\end{equation}

Similarly, for a dimer-resolved Stark detuning profile of $\Delta_j = jh$ where the first dimer defines the zero of the gradient ($\delta_1 = 0$),
\begin{equation}
    \begin{split}
    \partial_h \langle X_D \rangle \vert_{h=0} &= \frac{2\sqrt{2}q}{M (q^2 + 2)} \sum_j \beta_0 (j^2 -2j +1) \\
    &= \frac{\sqrt{2}q \beta_0}{3(q^2 + 2)}(2M^2 -3M +1),
    \end{split}
\end{equation}
giving:
\begin{equation}
    \begin{split}
        \frac{\lvert \partial_\theta \langle X_D \rangle \rvert^2}{\text{Var}(X_D)} &= \frac{2q^2 \beta_0^2}{9(q^2 + 2)^2} \cdot (4M^5 - 12M^4 + 13M^3 -6M^2 + M)\\
        &= \frac{8q^2 \beta_0^2}{9(q^2 + 2)^2} M^5 + \mathcal{O}(M^4).
    \end{split}
\end{equation}

\subsection{Linking dimer-local observable to implementable emitter-local observable}

Having established the QFI-scaling using the synthetic operator $X_D$, it can be shown that this maps directly to the implementable emitter-local operator string: the staggered collective-X,
\begin{equation}
    M_x^{\rm stag} := \frac{1}{N} \sum_{k=1}^N (-1)^k X_k = \frac{1}{M} \sum_{j=1}^{M} \mathcal{X}_j,
\end{equation}
where $\mathcal{X}_j$ is the staggered-X operator across dimer-$j$, $\mathcal{X}_j = (X_{2j} - X_{2j-1})/2$, and $N$ is even.

We first note that the staggered-X operator can be expressed as:
\begin{equation}
    \mathcal{X}_j = \frac{\zeta_j - \tau_j}{\sqrt{2}},
    \quad \zeta_j = \ketbra{ee}{\Psi^-}_j + \ketbra{\Psi^-}{ee}_j,
\end{equation}
where $\zeta_j$ couples the singlet to the bright-sector. Thus, from our operating point at $\delta_j = 0$, we arrive at:
\begin{equation}
    \partial_{h} \langle \mathcal{X}_j \rangle |_{0} = \frac{1}{\sqrt{2}}[\partial_{h} \langle \zeta_j \rangle |_{0} - \partial_{h} \langle \tau_j \rangle |_{0}],
\end{equation}
of which only the latter remains to be explicitly stated.

From Eq.~\eqref{app:eqn:G}, we already have the first-order response $\rho_1^{(1,1)}$ for a single dimer; and since for any observable $O$ on the dimer, $\partial_h \langle O \rangle |_{0} = \text{Tr}[O \rho_1^{(1,1)}]$, we arrive at:
\begin{equation}
    \begin{split}
        \partial_{h} \langle \zeta_j \rangle |_{0} = \text{Tr}[\zeta \rho_1^{(1,1)}]_j &= 2 \,\text{Re} \langle \Psi^- \lvert \rho_1^{(1,1)} \rvert ee \rangle \\
        &= -\frac{4\sqrt{2}q}{\gamma_R(q^2 + 2)^2},
    \end{split}
\end{equation}
and thus:
\begin{equation}
    \begin{split}
        \partial_{h} \langle \mathcal{X}_j \rangle |_{0} &= \frac{1}{\sqrt{2}} \left[ -\frac{4\sqrt{2}q}{\gamma_R(q^2 + 2)^2} - \frac{4\sqrt{2}q (2q^2 + 1)}{\gamma_R(q^2 + 2 )^2} \right] \\
        &= -\frac{8q(q^2+1)}{\gamma_R (q^2+2)^2}  =: g_{\mathcal{X}}.
    \end{split}
\end{equation}

Applying Eq.~\eqref{eqn: generic-proj-resp} in general then gives:
\begin{equation}
    G_{jk}^{(\mathcal X)} = \text{Tr}[\mathcal{X}_j \rho_M^{(1,k)}] =
    \begin{cases}
        0, & j<k, \\
        g_\mathcal X, & j=k, \\
        2 g_\mathcal X, & j>k.
    \end{cases}
\end{equation}

Now, let us define the $x_j := \partial_h \langle \mathcal{X}_j \rangle|_{h=0} = \text{Tr}[\mathcal{X}_j \rho^{(1)}]$. For uniform detuning perturbations of $\delta_j = h$, we see that $x_j = g_{\mathcal X} (2j-1)h$. Thus,
\begin{equation}
    \begin{split}
        \partial_h \langle M_x^{\rm stag} \rangle |_{h=0} &= \frac{1}{M} \sum_{j=1}^{M} \partial_h \langle \mathcal{X}_j \rangle|_{h=0} \\
        &= \frac{1}{M} \sum_{j=1}^{M}  g_{\mathcal{X}} (2j-1) =  g_{\mathcal{X}} M.
    \end{split}
\end{equation}

To obtain the variance of the observable, we first note that for a single dimer:
\begin{equation}
    \text{Var}_{\rho_D} (\mathcal X) = \frac{q^2 + 4}{2(q^2 + 2)},
\end{equation}
which then leads to:
\begin{equation}
    \text{Var}_{\rho_0} (\mathcal X) = \frac{1}{M} \text{Var}_{\rho_D} (\mathcal X) = \frac{1}{M} \cdot \frac{q^2 + 4}{2(q^2 + 2)}.
\end{equation}

The error-propagation quantity thus evaluates to:
\begin{equation}
    \frac{\lvert \partial_h \langle M_x^{\rm stag} \rangle \rvert^2}{\text{Var}(M_x^{\rm stag})} = \frac{128q^2 (q^2+1)^2}{\gamma_R^2(q^2+2)^3(q^2+4)} M^3.
\end{equation}

Intuitively, $M_x^{\rm stag}$ can thus track detuning changes because it contains information both on internal dark-sector rotations and bright-dark sector correlations.

Also for Stark gradient perturbations of $\delta_j = (j-1)h$, we see that $x_j = g_\mathcal{X} (j^2 -2j +1)h$, and we have:
\begin{equation}
    \begin{split}
        \partial_h \langle M_x^{\rm stag} \rangle |_{h=0} 
        &= \frac{1}{M} \sum_{j=1}^{M}  g_{\mathcal{X}} (j^2 -2j +1) \\ &= g_{\mathcal{X}} \left( \frac{1}{3}M^2 -\frac{1}{2}M + \frac{1}{6}  \right),
    \end{split}
\end{equation}
and thus:
\begin{equation}
    \frac{\lvert \partial_h \langle M_x^{\rm stag} \rangle \rvert^2}{\text{Var}(M_x^{\rm stag})} = \frac{g_{\mathcal{X}}}{9} M^5 + \mathcal{O}(M^4).
\end{equation}

\section{Measurement protocol with direct readout}
\label{app:CFI}

For the main observable $M_x^{\rm stag}$, we consider the case where we measure only the eigenvalue of $M_x^{\rm stag}$ in each shot. Let $x_j \in \{+1,-1\}$ denote the outcome of an $X_j$ measurement. A complete $X$-basis string is denoted $\mathbf{x} = (x_1, \cdots, x_N)$, and its staggered magnetization is:
\begin{equation}
    m(\mathbf{x}) = \frac{1}{N} \sum_{j=1}^{N}(-1)^j x_j.
    \label{eq: appG_m_of_x}
\end{equation}

The single-emitter projectors onto the $X_j$ eigenstates are:
\begin{equation}
    \Pi_{x_j}^{(j,X)} = \frac{\mathbb{I}+x_j X_j}{2},
\end{equation}
and hence the projective measurement of the complete $X$ string has POVM elements:
\begin{equation}
    \Pi_{\mathbf{x}}^{(X)} = \bigotimes_{j=1}^{N} \frac{\mathbb{I}+x_jX_j}{2},
    \quad \mathbf{x} \in \{\pm1\}^N.
    \label{eq: appG_XstringPOVM}
\end{equation}

If only $M_x^{\rm stag}$ is retained, all strings with the same value of Eq.~\eqref{eq: appG_m_of_x} are coarse-grained into a single outcome. The corresponding POVM is therefore:
\begin{equation}
    E_m^{\rm stag} = \sum_{\mathbf{x}: \, m(\mathbf{x})=m} \Pi_{\mathbf{x}}^{(X)},
    \label{eq: appG_MstagPOVM}
\end{equation}
where $m \in \{-1,-1+2/N,\ldots,1-2/N,1\}$. Equivalently, $E_m^{\rm stag}$ is the spectral projector of $M_x^{\rm stag}$ associated with eigenvalue $m$. 

We also note that if the only available projective readout is in the emitter Z-basis, it is sufficient to apply alternating $+\pi/2$ and $-\pi/2$ rotations about $Y$, and then measure the total $Z$-magnetisation / total excitation number. Repeated measurements of this collective population can be used to estimate $\langle M_x^{\rm stag} \rangle$ and $\text{Var}(M_x^{\rm stag})$, but cannot recover the full distribution $p_m (\theta)$ as that requires site-resolved bit-string readout.

\section{Waveguide-output readout by balanced full-record homodyne detection}
\label{app:homodyne}

This approach directly monitors the field leaving the downstream output port of the chiral waveguide. This is particularly natural for the present model because the same collective operator that mediates radiative coupling also determines the output field.

First, let $\rho_{ss}(\theta)$ denote the sensing steady state at parameter value $\theta$. Immediately before optical/microwave readout, we apply $Z$ operator on either the even or odd sites,
\begin{equation}
    U_Z = \prod_{j\in\mathbb{O}} Z_j,
    \quad
    \rho_{\rm ro}(\theta) = U_Z \rho_{ss}(\theta) U_Z^{\dagger},
    \label{eq: appG_Zpulse}
\end{equation}
where $\mathbb{O}$ denotes the chosen odd sites. This operation changes the relative phase structure that makes the prepared state dark with respect to the collective output and thereby converts part of the stored emitter coherence into a bright component. The system is then allowed to radiate for a readout window, $0\leq t\leq T_{\rm ro}$.

In the fully right-chiral limit, the monitored collapse operator is $L_R = \sum_{j=1}^{N}\sqrt{\gamma_{R,j}} \sigma_j^{-}$, which reduces to $L_R = \sqrt{\gamma_R}\sum_j\sigma_j^{-}$ for equal couplings. For vacuum input, input-output theory gives:
\begin{equation}
    b_{\rm out}(t) = b_{\rm in}(t) + L_R(t),
    \quad \langle b_{\rm in}(t) \rangle=0.
    \label{eq: appG_inputOutput}
\end{equation}
The local oscillator used for homodyne detection is introduced downstream of the emitter array and is therefore not part of $b_{\rm in}$ in Eq.~\eqref{eq: appG_inputOutput}.

Balanced homodyne detection mixes $b_{\rm out}$ with a strong local oscillator of phase $\phi$ on a 50:50 beam splitter and records the difference photocurrent between the two output detectors. The associated system quadrature is:
\begin{equation}
    X_\phi := e^{-i\phi} L_R + e^{i\phi} L_R^\dagger.
\end{equation}
To state the physical POVM explicitly, we partition the readout interval into $K = T_{\rm ro}/dt$ temporal bins and define the normalized output mode in bin $k$ by:
\begin{align}
    b_k &= \frac{1}{\sqrt{dt}} \int_{t_k}^{t_k+dt} b_{\rm out}(t)\,dt, \quad  [b_k,b_{\ell}^{\dagger}]=\delta_{k\ell}. \\
     Q_{k,\phi} &= \frac{1}{\sqrt2} (e^{-i\phi} b_k + e^{i\phi} b_k^\dagger).
\end{align}
Let $\ket{x_k;\phi}$ denote its generalised eigenstate, i.e. $Q_{k,\phi} \ket{x_k;\phi}=x_k\ket{x_k;\phi}$, such that $\int_{-\infty}^{\infty}dx_k\, \ket{x_k;\phi}\!\bra{x_k;\phi} = \mathbb{I}_k$. Ideal time-resolved homodyne detection then has the output-field POVM:
\begin{equation}
    \Pi_{\mathbf{x}}^{(hom,\phi)} d^K \mathbf{x} =\bigotimes_{k=1}^{K} \ketbra{x_k;\phi}{x_k;\phi}\,dx_k.
    \label{eq: appG_homodynePOVM}
\end{equation}
Equivalently, writing $dY_k = \sqrt{2dt}\,x_k$, one experimental shot produces the complete record $\mathbf{Y} = (dY_1,\cdots,dY_K)$. Note that vacuum shot noise has variance $\text{Var}(dY_k) = dt$. Detector efficiency, $0\leq\eta_d\leq1$, can be represented by a beam-splitter loss channel of transmissivity $\eta_d$ preceding the ideal homodyne POVM.

Let $\mathcal{Y}_t=\{dY_s:0\leq s<t\}$ denote the homodyne record acquired up to time $t$. We denote by:
\begin{equation}
    \rho_c(t;\theta) := \rho\!\left(t\,\middle|\,\mathcal{Y}_t,\theta\right),
\end{equation}
the emitter density operator conditioned on this measurement record, with initial condition
$\rho_c(0;\theta)=\rho_{\rm ro}(\theta)$. Thus, $\rho_c(t;\theta)$ is a stochastic state whose evolution depends on the homodyne outcomes obtained at all earlier times.

Under homodyne monitoring, the conditional state evolves according to:
\begin{equation}
    d\rho_c = \mathcal{L}_{\rm ro}[\rho_c]\,dt + \sqrt{\eta_d} \mathcal{H}[e^{-i\phi}L_R]\rho_c \,dW_t,
\end{equation}
where $\mathcal{L}_{\rm ro}$ is the unconditional Liouvillian during the readout window, and $\mathcal{H}[A]\rho := A\rho+\rho A^\dagger - \text{Tr}\!\left[(A+A^\dagger)\rho\right]\rho$.

The observed homodyne increment thus obeys:
\begin{equation}
    dY_t = \sqrt{\eta_d}\, \langle X_\phi\rangle_{c,t} dt + dW_t,
    \quad \langle X_\phi\rangle_{c,t} = \text{Tr} [X_\phi \rho_c(t)].
    \label{eq: appG_homodyneCurrent}
\end{equation}
where $dW_t$ is a Wiener increment satisfying $\mathbb{E}[dW_t]=0$ and $dW_t^2=dt$.
Since $\rho_c(t;\theta)$ depends on all earlier measurement outcomes, the homodyne increments are generally not independent random variables, but are conditionally Gaussian given the preceding record.

For a time-discretised record $\mathbf{Y}$, let $\rho^c_{\theta,k}$ denote the conditional state immediately before the $k$-th increment such that:
\begin{equation}
    \rho^c_{\theta,k} := \rho_c(t_k;\theta) =\rho\!\left(t_k\,\middle|\,\mathbf{Y}_{<k},\theta\right),
    \quad
    \mathbf{Y}_{<k}=(dY_1,\ldots,dY_{k-1}),
\end{equation}
and we define: $x_{\theta,k} := \text{Tr}[X_\phi \rho^c_{\theta,k}]$. Then we have:
\begin{equation}
    p(dY_k \vert \mathbf{Y}_{<k}, \theta) = \frac{1}{\sqrt{2\pi dt}} \exp \left[-\frac{(dY_k-\sqrt{\eta_d} x_{\theta,k} dt)^2}{2dt} \right].
    \label{eq: appG_conditionalGaussian}
\end{equation}
By the chain rule of probability, the likelihood of the complete record is:
\begin{equation}
    p (\mathbf{Y} \mid \theta) = \prod_{k=1}^{K} p( dY_k \mid \mathbf{Y}_{<k},\theta).
    \label{eq: appG_recordLikelihood}
\end{equation}
Up to a $\theta$-independent normalization constant, the corresponding log likelihood is:
\begin{equation}
    \ln p(\mathbf{Y}\mid\theta) = -\frac{1}{2}\sum_{k=1}^{K} \frac{[dY_k - \sqrt{\eta_d} x_{\theta,k} dt]^2}{dt} + \text{const.}
    \label{eq: appG_recordLogLikelihood}
\end{equation}

One point of note is that experimentally, it is likely that one lacks two independent ports for both coherent driving and for implementing rotations. As a result, it is convenient to turn off the coherent drive ($\Omega_j = 0 \, \forall j$) at the start of the readout protocol until $T_{\rm ro}$.

\section{Preparation time scaling}
\label{app:time}

\subsection{Spectral properties at $\delta_j = 0$, $\Omega_j < \gamma_R/2$}

Here we analytically prove that if we further levy $\delta_j = 0$ and $\Omega_j < \gamma_R/2$, we obtain a constant spectral gap together with a Jordan-chain structure at the spectral-gap-eigenvector-manifold. It behaves as a metastable manifold that decouples spectral gap scaling from preparation time scaling.

\subsubsection{Constant spectral gap value of $\lambda_\star = -\gamma_R/2$}

A consequence of the matrix having a block-triangular structure is that the spectrum of the entire matrix is just the union (with algebraic multiplicity allowed) of the spectra of its block-diagonals, i.e. $\text{spec}(\mathcal{L}) = \text{spec}(\mathcal{L}^{\#}) = \uplus_{m=0}^{N} \text{spec}(B_m)$. This allows for quick numeric validation of the spectra without running into inaccuracy due to the large condition-number of the full Liouvillian.
We can also argue by contradiction that the spectral gap $\lambda_\star = -\gamma_R/2$ cannot decrease in norm as $N$ increases. Noting that the implication of Heisenberg-block triangularity on the spectra can computed by the characteristic polynomial $\chi_N (z)$ given by:
\begin{equation}
    \chi_N(z)=z\prod_{m=1}^N\beta_m(z),
    \quad \beta_m(z):=\det(z-B_m),
    \label{eq:support_factorization}
\end{equation}
we see it is sufficient to show that every support layer $B_m$ has spectral bound $-\gamma_R/2$.

To begin, we temporarily return to the Schr\"odinger picture and consider the first two emitters in the chain. At resonance, they possess the exact output-dark state, $\ket{D} \propto \frac{i\gamma_R}{\Omega} \ket{gg} - \ket{ge} + \ket{eg}$, which is annihilated by both the two-emitter effective non-Hermitian Hamiltonian $H_{\rm eff}^{(2)}$ and the collective output operator: $c=\sigma_1^-+\sigma_2^-$.
Writing the projectors:
\begin{equation}
    p = \frac{\ketbra{D}{D}}{\langle D|D\rangle},
    \quad p^{\perp} = \mathbb{I} - p,
\end{equation}
we note that the Schr\"odinger-Liouvillian $\mathcal{L}_m$ is a linear map on the bounded operator space $\mathcal{B}(\mathcal{H}_{1:m}) \sim \mathcal{B}$, and that projectors $p$ and $q$ only act non-trivially on $\mathcal{H}_{1,2}$. Thus, the full $m$-emitter operator space can be split into four sectors: $(p\mathcal Bp , q\mathcal Bp , p\mathcal Bq , q\mathcal Bq)$. In a basis ordered according to these sectors, the Schr\"odinger-Liouvillian can also be arranged in a block-triangular manner:
\begin{equation}
    \mathcal{L}_m \sim 
    \begin{pmatrix}
        \mathcal{L}_{m-2} & \bullet & \bullet & \bullet \\
        0 & K_m^+ & 0 & \bullet \\
        0 & 0 & K_m^- & \bullet \\
        0 & 0 & 0 & M_m
    \end{pmatrix}
\end{equation}
owing to the above annihilation conditions of the first dimer; the bullet points $\bullet$ indicate blocks with values irrelevant to our present discussion. This Schr\"odinger-block triangularity now imposes the spectral condition:
\begin{equation}
    \chi_m (z) = \chi_{m-2} (z) \cdot \kappa_m^+ (z) \cdot \kappa_m^- (z) \cdot \pi_m (z),
    \label{eq:dark_cell_factorisation_gap}
\end{equation}
where we have defined the characteristic polynomials, $\pi_m(z)=\det(z-M_m)$ and $\kappa_m^\pm (z) = \det(z-K_m^\pm)$.

The gist of the proof hereon out is to couple the spectral conditions from the Schr\"odinger and Heisenberg pictures to show that:
\begin{equation}
    \beta_m (z) = \beta_{m-1}(z) \cdot \pi_m (z),
\end{equation}
or equivalently,
\begin{equation}
    \text{spec}(B_m) = \text{spec}(B_{m-1}) \uplus \text{spec}(M_m),
    \label{eq: constant_gap_core}
\end{equation}
where $\uplus$ is the union with multiplicities/duplicate entries retained, and to then show that $\pi_m(z)$ (or equivalently $\text{spec}[M_m]$ have eigenvalues all more negative than $\lambda_\star$.

To start, we find that the two dark-bright coherence blocks $K_m^\pm$ inherit exactly the preceding rightmost-support spectrum,
\begin{equation}
    \det(z-K_m^+) = \det(z-K_m^-) = \beta_{m-1}(z).
    \label{eq:coherence_inheritance}
\end{equation}
To see why, first observe that when we compare the $m$-emitter vectorised Liouvillian $\hat{\mathcal{L}}_m$ with the $2m$-emitter effective Hamiltonian split after the first $m$-sites:
\begin{align}
    \hat{\mathcal{L}}_m &= H_{\rm eff}^{(m)} \otimes \mathbb{I} + \mathbb{I} \otimes H_{\rm eff}^{(m)*} + \gamma_R c_m \otimes c_m, \\
    H_{\rm eff}^{(2m)} &= H_{\rm eff}^{(m)} \otimes \mathbb{I} + \mathbb{I} \otimes H_{\rm eff}^{(m)} - \gamma_R c_m \otimes c_m^\dagger,
\end{align}
they have similar forms that are interconvertible by acting on the second half of the bipartition with a partial transpose $T_B$ and unitary transform:
\begin{equation}
    (\mathbb{I} \otimes U_n) [H_{\rm eff}^{(2n)}]^{T_B} (\mathbb{I} \otimes U_n^{-1}) = \hat{\mathcal L}_n,
\end{equation}
where $U_n = Z^{\otimes n} R_n$ being a composition of an all-chain Pauli-Z and a permutation operator $R_n$ that reverses the order of the $n$-sites. Therefore, $\text{det}(z-[H_{\rm eff}^{(2n)}]^{T_B}) = \text{det}(z- \hat{\mathcal{L}}_n) = \chi_n (z)$.
We find, after a bit of algebra, that the partial transpose preserves the spectrum of the effective Hamiltonian:
\begin{equation}
    \text{det}(z-[H_{\rm eff}^{(2n)}]^{T_B}) = \text{det}(z-H_{\rm eff}^{(2n)}).
    \label{eq:Heff_transpose}
\end{equation}
From this relation, $\hat{\mathcal{L}}_m$ and $H_{\rm eff}^{(2n)}$ have the same spectrum:
\begin{equation}
    \text{det}(z - H_{\rm eff}^{(2n)}) = \chi_n (z)
\end{equation}
and an extension gives rise to:
\begin{equation}
    \text{det}(z - K_m^\pm) = \frac{\text{det}(z - H_{\rm eff}^{(2n-2)})}{\text{det}(z - H_{\rm eff}^{(2n-4)})} = \frac{\chi_{n-1}}{\chi_{n-2}} = \beta_{m-1}.
    \label{eq:Heff_spectrum}
\end{equation}

Next, we find that the bright-sector block $M_m$ has a spectral bound obeying:
\begin{equation}
    s(M_m) := \max_{\lambda\in\text{spec}(M_m)} \text{Re}[\lambda] \leq s(M_2) < -\frac{\gamma_R}{2}.
    \label{eq:killed_bound}
\end{equation}
To show this, we prove that the semigroup generated by $M_m$ (i.e. $e^{tM_m}$) is trace non-increasing while still being completely positive. The trace non-increasing property is intuitive: probability amplitude can leave the bright first-dimer subspace, but strict chirality prevents the downstream chain from feeding it back upstream, i.e. defining the projector $Q_m := p^{\perp} \otimes \mathbb{I}_{3\cdots m}$, we have $Q_m \mathcal{L}_m Q_m =0$. Complete positivity is a consequence of $e^{tM_m} = Q_m e^{t\mathcal{L}_m} Q_m$. To show this, we let $Y(t) = Q_m \rho(t) Q_m$ and consider: 
\begin{equation}
    \begin{split}
        \frac{dY}{dt} &= \frac{d}{dt} [Q_m \rho(t) Q_m]\\ 
        &= Q_m \mathcal{L}_m [\rho(t)] Q_m \\
        &=  Q_m \mathcal{L}_m [P_m \rho(t) P_m + P_m \rho(t) Q_m + Q_m \rho(t) P_m + Q_m \rho(t) Q_m] Q_m \\
        &= Q_m \mathcal{L}_m[Q_m \rho(t) Q_m] Q_m \\
        &= M_m [Y(t)],
    \end{split}
\end{equation}
where the non-bright sectors are eliminated due to dark-sector kets/bras being annihilated. We have both $Y(t) = e^{tM_m} [Y(0)]$ and $Y(t) = Q_m e^{t \mathcal{L}_m}[Y(0)] Q_m$, so by uniqueness of ODEs, the relation $e^{tM_m} = Q_m e^{t\mathcal{L}_m} Q_m$ is fulfilled.

The Perron-Frobenius theorem~\cite{berman1994nonnegative} thus applies to the matrix-representation of superoperator $e^{tM_m}$, so its spectral bound $s_m=s(M_m)$ corresponds to a completely positive eigenvector which we can reshape back to obtain the eigenoperator $X_m \geq 0$. Now we trace out all sites downstream of the first dimer, defining:
\begin{equation}
    x_2:=\text{Tr}_{3,\cdots,m}(X_m).
\end{equation}
Because $X_m\geq0$ and $X_m\neq0$, we have $x_2\geq0$ and $\text{Tr}(x_2)=\text{Tr}(X_m)>0$, so $x_2\neq0$. Moreover, $X_m=qX_mq$ implies $qx_2q=x_2$.

Strict chirality levies the relation: $\text{Tr}_{3,\ldots,m}\!\left[\mathcal{L}_m(X)\right] = \mathcal{L}_2\!\left[\text{Tr}_{3,\ldots,m}(X)\right]$, so substituting terms and projecting the first dimer back into $q$ thus gives:
\begin{equation}
    M_2(x_2)=q\mathcal{L}_2(x_2)q = \text{Tr}_{3,\ldots,m}\!\left[M_m(X_m)\right]=s_mx_2.
\end{equation}
Thus, $s_m$ is itself an eigenvalue of the isolated two-emitter problem, which immediately implies:
\begin{equation}
    s(M_m)\leq s(M_2).
\end{equation}
Direct evaluation of $M_2$ finally gives $s(M_2)<-\frac{\gamma_R}{2}$ for $0<\Omega/\gamma_R<1/2$, which establishes Eq.~\eqref{eq:killed_bound}.

We can now derive the spectral recursion for the Heisenberg support layers. Inserting Eq.~\eqref{eq:coherence_inheritance} into Eq.~\eqref{eq:dark_cell_factorisation_gap} gives:
\begin{equation}
    \chi_m(z) = \chi_{m-2}(z) \cdot \beta_{m-1}^2(z) \cdot \pi_m(z).
\end{equation}
Meanwhile, the Heisenberg support factorization at size $m-1$ gives $\chi_{m-1}(z) = \chi_{m-2}(z) \cdot \beta_{m-1}(z)$. Since $\chi_m(z) = \chi_{m-1}(z) \cdot \beta_m(z)$, comparison yields:
\begin{equation}
    \beta_m(z)=\beta_{m-1}(z) \cdot \pi_m(z),\quad m\geq2.
    \label{eq:beta_recursion}
\end{equation}
Equivalently, including algebraic multiplicity:
\begin{equation}
    \text{spec}(B_m)=\text{spec}(B_{m-1})\uplus\text{spec}(M_m).
    \label{eq:B_spectrum_recursion_gap}
\end{equation}
This identity captures the physical content of the entire construction: each newly allowed rightmost-support layer contains one inherited copy of the preceding layer, together with a new bright sector whose modes all decay faster than $\gamma_R/2$.

Now the proof by contradiction appears simply. Suppose there exists some $B_m$ containing an eigenvalue $\lambda$ with $\text{Re}[\lambda] > -\gamma_R/2$, and we choose the minimum such $m$ if more than one exists. From~\eqref{eq: constant_gap_core}, any such hypothetical slow root of $B_m$ must come from either $B_{m-1}$ or $M_m$. 

If it came from $B_{m-1}$, then this contradicts minimality of $m$. As a base case, the spectrum of $B_1$ is $\{-\frac{\gamma_R}{2} , -\frac{3 \gamma_R}{4} \pm \frac{1}{4} \sqrt{\gamma_R^2 - 16 \Omega^2} \}$, so there is no such eigenvalue at $m=1$. But if it came from $M_m$, then this contradicts~\eqref{eq:killed_bound}. Therefore, no such slower eigenvalue can exist and the spectral gap $\lambda_\star$ must be constant.

Additionally, since $\lambda_\star$ is present in $B_1$ and is inherited once by every subsequent $B_m$, $\lambda_\star$ is an eigenvalue of algebraic multiplicity-$N$ of the full Liouvillian. The following subsection turns to the Jordan structure associated with this repeated gap eigenvalue and to its dynamical implications.

\subsubsection{Jordan-chain structure of eigenvector-manifold associated to spectral gap}

Moving back to the Schr\"odinger picture for convenience, we now determine the Jordan structure associated with the repeated gap eigenvalue $\lambda_\star$. 
For $M=N/2$ number of dimers with the corresponding effective Hamiltonian by $H_{\rm eff}^{(M)}$,  relevant eigenvalue is $\epsilon_\star= i\lambda_\star = -i\gamma_R/2$.
For a single dimer at zero-detuning, with $r:=\gamma_R/\Omega$ and the two-site effective Hamiltonian $H_{\rm eff}^{(1)}$, two useful normalized right states are:
\begin{align}
    \ket{D} &= \frac{i r\ket{gg}-\ket{ge}+\ket{eg}}{\sqrt{r^2+2}}, \\
    \ket{b} &= \frac{-\ket{gg}+i r\ket{ge}+\ket{ee}}{\sqrt{r^2+2}},
\end{align}
which obey,
\begin{align}
    H_{\rm eff}^{(1)}\ket{D}&=0, & C\ket{D}&=0, \\
    H_{\rm eff}^{(1)}\ket{b}&=\epsilon_\star\ket{b}, & \braket{D|b}&=0,
    \label{eq:right_local_relations}
\end{align}
where $C=\sigma_1^-+\sigma_2^-$. Noting that $\ket{b} = \frac{1}{\sqrt{2}} (\ket{T} - \ket{T^*}) \propto -\ket{gg} + \frac{i\gamma_R}{\Omega} \ket{ge} + \ket{ee}$, 2e can thus regard $\ket{b}$ as a `defect'-state leaking probability amplitude from the dark-sector into the bright-sector, while $\ket{D}$ is just the dark-dimer state.

Then for $M = N/2$ dimers, let us define for $j \in \{1, \cdots, M\}$ the undressed defect-position states:
\begin{equation}
    \ket{j} = \ket{D}^{\otimes(j-1)} \otimes \ket{b}_j \otimes \ket{D}^{\otimes (M-j)}.
\end{equation}
Unlike the dark product state, these states are not all exact eigenvectors of $H_{\rm eff}^{(M)}$: if $j<M$, the defect on dimer $j$ can radiate into downstream dimers. Only the terminal state is immediately an exact right eigenvector,
\begin{equation}
    H_{\rm eff}^{(M)}\ket M=\epsilon_\star\ket M.
    \label{eq:terminal_defect_eigenvector}
\end{equation}
The physical picture is nevertheless the same as in the bare defect basis: chirality allows $\ket{b}_j $ of the defect to move only downstream.

To make this statement precise, let us append an upstream dark dimer through:
\begin{equation}
    E_M\ket\psi_{2\cdots M}:=\ket D_1\otimes\ket\psi_{2\cdots M}.
\end{equation}
Since $H_{\rm eff}^{(1)}\ket D=0$ and $C\ket D=0$, every cascade term sourced by the first dimer vanishes, while strict chirality prevents downstream dimers from feeding back into it. Hence,
\begin{equation}
    H_{\rm eff}^{(M)} E_M=E_M H_{M-1}
    \label{eq:dark_embedding_intertwiner}
\end{equation}
Thus, every Jordan chain present for $M-1$ dimers is inherited exactly by the $M$-dimer problem inside the invariant subspace:
\begin{equation}
    \mathcal S_M:= \text{span}\{\ket{D}_1\}\otimes\mathcal H_{2\ldots M}.
\end{equation}

Appending one dimer therefore leaves only one question: whether the new copy of $\epsilon_\star$ forms a separate eigenvector, or whether it joins the inherited block to make the Jordan chain one site longer.

We proceed via induction. Suppose first that the $(M-1)$-dimer problem contains one Jordan block of length $M-1$ at $\epsilon_\star$. Eq.~\eqref{eq:dark_embedding_intertwiner} then guarantees that this entire block is inherited by the $M$-dimer problem. We show that the one new algebraic copy of $\epsilon_\star$ joins this inherited block rather than forming a separate eigenvector.

We first verify that exactly one new copy appears. A convenient left eigenbra of the single-dimer problem is:
\begin{equation}
    \bra{\widetilde b} :=\frac{\sqrt{r^2+2}}{2} \left(-\bra{gg}+ir\bra{eg}+\bra{ee}\right),
\end{equation}
which satisfies,
\begin{equation}
    \bra{\widetilde b}H_{\rm eff}^{(1)}=\epsilon_\star\bra{\widetilde b},
    \quad \braket{\widetilde b|b}=1,
    \quad \braket{\widetilde b|D}=0.
\end{equation}
Since we also have $\bra{D} H_{\rm eff}^{(1)}=0$ and $\bra{D} C^\dagger=0$, the product bra:
\begin{equation}
    \bra{\Lambda_M} := \bra{\widetilde b_1}\otimes\bra{D}_2\otimes\cdots\otimes\bra{D}_M
    \label{eq:left_upstream_gap_mode}
\end{equation}
is an exact left eigenvector of $H_{\rm eff}^{(M)}$ at $\epsilon_\star$. Importantly, $\bra{\Lambda_M}$ annihilates the inherited subspace $\mathcal{S}_M$, since its first-dimer factor is orthogonal to $\ket{D}$. Thus, appending the new upstream dimer introduces at least one additional algebraic copy of $\epsilon_\star$.

Letting $a_M$ denote the algebraic multiplicity of $\epsilon_\star$ in $H_{\rm eff}^{(M)}$, this gives $a_M\geq a_{M-1}+1$, and therefore $a_M\geq M$. On the other hand, every generalised eigenvector of $H_{\rm eff}^{(M)}$ at $\epsilon_\star$ produces one ket-dark and one bra-dark generalised eigenmode of the Liouvillian at $\lambda_\star$. Since the preceding spectral recursion already showed that $\lambda_\star$ has total algebraic multiplicity $2M=N$, we must have $a_M\leq M$. Hence, $a_M=M$; adding one upstream dimer introduces exactly one new generalised-eigenvector direction.

The final step is to show that this new direction extends the old Jordan chain. Let $\ket{\eta_M}$ be chosen such that:
\begin{equation}
    \braket{\Lambda_M|\eta_M}=1.
\end{equation}
Because the inherited subspace already contains all the other generalized directions at $\epsilon_\star$, its residual under $H_{\rm eff}^{(M)}-\epsilon_\star$ lies inside $\mathcal S_M$:
\begin{equation}
    \ket{f_M}:=(H_{\rm eff}^{(M)}-\epsilon_\star)\ket{\eta_M}\in\mathcal S_M.
\end{equation}
Physically, $\ket{f_M}$ is the dressed defect-transfer vector: it is what remains of the new upstream defect after all faster components outside the gap manifold are accounted for. To test whether this transfer actually joins the inherited Jordan block, we take the left gap eigenbra of that block,
\begin{equation}
    \bra{\Xi_M}:= \bra{D}_1\otimes\bra{\widetilde{b}_2}\otimes\bra{D}_3\otimes\cdots\otimes\bra{D}_M,
\end{equation}
and define the amplitude:
\begin{equation}
    \mu_M:=\braket{\Xi_M|f_M} =\bra{\Xi_M}(H_{\rm eff}^{(M)}-\epsilon_\star)\ket{\eta_M}.
    \label{eq:chain_merging_scalar}
\end{equation}
For a single inherited Jordan block, the range of $H_{\rm eff}^{(M)}-\epsilon_\star$ inside $\mathcal{S}_M$ is precisely the codimension-one subspace annihilated by $\bra{\Xi_M}$. Hence, if $\mu_M=0$, one could add a vector from $\mathcal{S}_M$ to $\ket{\eta_M}$ and turn it into a second wholly independent eigenvector. Conversely, $\mu_M\neq0$ means that such a cancellation is impossible, so the new direction must join the inherited chain; and we can show that this is the case.

Suppose we contract all dimers downstream of the first two against the dark-bra product state, then we define the partial-contraction linear map:
\begin{equation}
    R_M:= \mathbb I_{12}\otimes\bra{D}_3\otimes\cdots\otimes\bra{D}_M.
\end{equation}
such that $\bra{\Xi_M} = (\bra{D}_1\otimes\bra{\widetilde{b}_2}) R_M$. Since $\bra{D} H_{\rm eff}^{(1)}=0$ and $\bra{D} C^\dagger=0$, every term involving dimers $3,\cdots,M$ vanishes under this contraction, and therefore:
\begin{equation}
    R_M H_{\rm eff}^{(M)} = H_2 R_M.
    \label{eq:left_dark_contraction_intertwiner}
\end{equation}
This implies that the dressed transfer that joins the Jordan block is the same for every system size, $\mu_M=\mu_2$, and it suffices to show that $\mu_2 \neq 0$. By direct calculation, we obtain:
\begin{equation}
    \mu_2 = i\frac{\gamma_R^3(\gamma_R^2+2\Omega^2)}{2\Omega^2(\gamma_R^2-4\Omega^2)}.
    \label{eq:exact_chain_merging_amplitude}
\end{equation}
For the regime considered here, $0<\Omega/\gamma_R<1/2$, this quantity is non-zero. Hence, every newly added dimer extends the inherited Jordan chain by one. Starting from the simple single-dimer eigenvalue, induction gives:
\begin{equation}
    H_{\rm eff}^{(M)}\big|_{\mathcal{E}_\star} \sim J_M(\epsilon_\star),
    \quad M=N/2,
    \label{eq:heff_single_gap_jordan_block}
\end{equation}
where $\mathcal{E}_\star := \text{ker}(H_{\rm eff}^{(M)}-\epsilon_{\star} \mathbb{I})^M$ refers to the generalised-eigenmanifold spanned by the spectal gap-vectors, and $J_M(x)$ refers to a Jordan chain of length-$M$ at eigenvalue $x$.

Using this, we can now show that the Liouvillian hosts two parallel Jordan-chains in the spectral gap-associated eigenvector manifold. The gist is that one can attach the $\ket{\phi_r}$ on either the ket- or bra-side and complement it with the global dark state on the dual side, and either configuration will inherit the Jordan chain from $H_{\rm eff}$.

We let the dark steady-state ket be $\ket{D_M} := \ket{D}^{\otimes M}$, and we note that $c\ket{D_M} = 0$ and $H_{\rm eff} \ket{D_M} = 0$. It follows that for any arbitrary ket $\ket{\psi}$, we have $\gamma c \ketbra{\psi}{D_M} c^{\dagger}$ vanishing as well. Hence, we obtain:
\begin{align}
    \mathcal{L}(\ketbra{\psi}{D_M}) &= -i H_{M} \ketbra{\psi}{D_M}, \\
    \mathcal{L}(\ketbra{D_M}{\psi}) &= i \ketbra{D_M}{\psi} H_{M}^\dagger.
\end{align}

Now we choose a canonical generalized chain $\{\ket{\phi_r}\}_{r=1}^M$ for $H_{\rm eff}^{(M)}$ such that:
\begin{align}
    (H_{\rm eff}^{(M)}-\epsilon_\star)\ket{\phi_1}&=0, \\
    (H_{\rm eff}^{(M)}-\epsilon_\star)\ket{\phi_{r+1}}&=\ket{\phi_r}.
    \label{eq:canonical_heff_chain}
\end{align}
The terminal defect-position state $\ket{M}$ may be chosen as $\ket{\phi_1}$. Note that this canonical normalization is used only to state the algebraic Jordan relations. In Appendix~\ref{app:moments}, a physical transfer rate is restored when estimating the corresponding dynamics.

We can define $R_r = \ketbra{\phi_r}{D_M}$ and use the Jordan relations on $\ket{\phi_r}$ to get:
\begin{align}
    (\mathcal{L} - \lambda_\star) R_1 &= 0, \\
    (\mathcal{L} - \lambda_\star) R_{r+1} &= -iR_r,
\end{align}
which gives us the new Jordan-chain relations:
\begin{equation}
    (\mathcal{L} - \lambda_\star) \widetilde{R}_{r+1} = \widetilde{R}_r,
\end{equation}
where $\widetilde{R}_r = i^{r-1} R_r$.

Taking Hermitian adjoints, we arrive at a second and independent Jordan-chain where $L_r = \ketbra{D_M}{\phi_r}$ that is related by:
\begin{equation}
    (\mathcal{L} - \lambda_\star) \widetilde{L}_{r+1} = \widetilde{L}_r,
\end{equation}
where $\widetilde{L}_r = -i^{r-1} L_r$. Thus, the eigenvectors at $\lambda_\star$ are described by $J_M(-\gamma_R/2) \oplus J_M(-\gamma_R/2)$ with a total algebraic multiplicity of $M+M = N$.

An exact description of the Jordan-chain eigenvectors is cumbersome, but the physical meaning is simple. The terminal defect-position state $\ket{M}$ is the true right-eigenvector of the Jordan chain, while moving the leading edge upstream produces generalized eigenvectors dressed by faster downstream components. In this sense, the Jordan chain describes the successive purification of dimers from upstream to downstream. The dynamical consequence of this structure is that preparation-time can be taken linear-in-$N$; this topic is discussed separately in Appendix~\ref{app:moments}.

\subsection{Preparation time upper bound based on spectral-gap Jordan chain}
\label{app:moments}

Much progress has recently been made on finding semigroup bounds for Liouvillian relaxation to steady-state; however, these methods require the steady-state to be invertible (i.e. unique and full-rank), which is not the case for our model with amplitude-damping-like noise. Here, we instead use the Jordan-chain structure derived in Appendix~\ref{app:Heisenberg} to obtain an upper bound for preparation time scaling despite a constant spectral gap. The only assumption we make in this argument is that late-time dynamics is controlled by the spectral-gap Jordan manifold, and that the faster manifolds do not generate an additional preparation-time scale that grows parametrically with $N$. This is corroborated with numerical tests.

We first define the reduced Liouvillian $\widetilde{\mathcal L}$ on the trace-zero subspace, i.e. with the steady-state eigenmode removed. Since a non-normal propagator need not decrease monotonically in norm, it is convenient to define the worst-case relaxation time as:
\begin{equation}
    T_\epsilon:=\inf\left\{t_0>0:\sup_{s\ge t_0}\left\|e^{\widetilde{\mathcal L}s}\right\|_{2\rightarrow2}\le\epsilon\right\}.
    \label{eq:ansatz_tail_time}
\end{equation}
This upper-bounds the preparation time from any initial state. To see this, consider for some time $s$ a state's difference to steady-state as,
\begin{equation}
    \rho(s) - \rho_{ss} = e^{\mathcal{L}s} \left[ \rho(0) -\rho_{ss} \right],
\end{equation}
for any arbitrary initial-state at $s=0$. Then it follows that,
\begin{equation}
    \norm{\rho(s) - \rho_{ss}}_2 \leq \norm{ e^{\mathcal{L}s}}_{2 \rightarrow 2} \norm{\rho(0) - \rho_{ss}}_2 \leq \sqrt{2} \norm{ e^{\mathcal{L}s}}_{2 \rightarrow 2}.
\end{equation}

Hence for every $s > T_{\epsilon}$,
\begin{equation}
    \norm{\rho(s) - \rho_{ss}}_2 \leq \sqrt{2} \epsilon.
\end{equation}

Explicitly, we can also show that it upper-bounds trace-distance. Recall that trace distance is based on the 1-norm: $D(\rho, \sigma) = \frac{1}{2} \norm{\rho - \sigma}_1$. For a $d$-dimensional Hilbert space, it holds that $\norm{X}_1 \leq \sqrt{\text{ran } X} \norm{X}_2 \leq \sqrt{d} \norm{X}_2$, so the relation becomes:
\begin{equation}
    D(\rho(s), \rho_{ss}) \leq \frac{\sqrt{d}}{2} \norm{\rho(s) - \rho_{ss}}_2 \leq \sqrt{\frac{d}{2}} \epsilon.
\end{equation}

We now proceed with the late-time ansatz. Appendix~\ref{app:Heisenberg} shows that for even $N$, $\delta_j=0$, and $0<\Omega/\gamma_R<1/2$, the spectral-gap manifold consists of two Jordan chains of length $M = N/2$ with spectral gap $\Delta = |\lambda_\star| = \gamma_R/2$. From the assumption that only the spectral-gap eigenmanifold is dynamically relevant and all other faster modes relax over a pre-asymptotic time $t_{pre}=\mathcal O(\gamma_R^{-1})$ which does not grow parametrically with $N$, the propagator can thus be modeled by the longest Jordan chain:
\begin{equation}
    \norm{e^{\tilde{\mathcal{L}}t}}_{2 \rightarrow 2} \approx e^{-\Delta t} \sum_{k=0}^{M -1} \frac{t^k}{k!}  \norm{S_M^k}_{2 \rightarrow 2}.
\end{equation}

To determine $\norm{S^k}_{2 \rightarrow 2}$, it is necessary to determine the physical rate at which states evolve down the layers of the Jordan chain. Let us work at the level of the effective no-jump non-Hermitian Hamiltonian, defining:
\begin{equation}
    A_M:=H_{\rm eff}^{(M)}-\epsilon_{\star},
    \quad \mathcal{K}_r:=\ker A_M^r,
    \quad r=1,\ldots,M.
\end{equation}
Since Appendix~\ref{app:Heisenberg} establishes two single length-$M$ blocks, we choose one of thsee chains and have $\dim\mathcal K_r=r$. We may therefore choose a physical orthonormal basis $\{\ket{u_r}\}_{r=1}^M$ with:
\begin{equation}
    \ket{u_r}\in\mathcal K_r\cap\mathcal K_{r-1}^{\perp},
    \quad \mathcal K_0:=\{0\},
\end{equation}
where physically $u_r$ is the new direction that emerges when the generalised eigenspace is enlarged upon shifting up from level $r-1$ to level $r$ of the Jordan chain.
In this basis, the effective Hamiltonian restricted to the spectral gap-sector $\mathcal{E}_\star = \text{ker}(H_{\rm eff}^{(M)}-\epsilon_{\star} \mathbb{I})^M$ is
\begin{equation}
    \left.(-iH_{\rm eff}^{(M)})\right|_{\mathcal{E}_\star} = - \Delta \, \mathbb{I} + S_M,
    \label{eq:physical_gap_nilpotent}
\end{equation}
where $S_M=-iA_M|_{\mathcal{E}_\star}$ is strictly upper triangular and carries the physical dimensions of a rate.

The successive-layer transfer rates can now be fixed directly. The effective no-jump Hamiltonian obeys the relation:
\begin{equation}
    (-iH_{\rm eff}^{(M)})+(-iH_{\rm eff}^{(M)})^\dagger=-\gamma_R c_M^\dagger c_M,
\end{equation}
with $c_M = \sum_{m=1}^M \sigma_m^-$ being the collective output operator as usual. Restricting this to the spectral gap-sector and using Eq.~\eqref{eq:physical_gap_nilpotent} gives us:
\begin{equation}
    S_M + S_M^\dagger = \gamma_R \left( \mathbb{I} - P_\star c_M^\dagger c_M P_\star \right),
    \label{eq:physical_gap_dissipativity}
\end{equation}
where $P_\star$ is the projector onto the spectral gap-sector. Because $S_M$ has zero diagonal, Eq.~\eqref{eq:physical_gap_dissipativity} first implies $\|c_Mu_r\|=1$ for every $r \in \{1, \cdots, M-1\}$. Then, for the single-layer transition, we have:
\begin{equation}
    g_r:=\left|(S_M)_{r,r+1}\right| =\gamma_R \left| \langle u_r| c_M^\dagger c_M |u_{r+1} \rangle \right| \le \gamma_R,
    \label{eq:physical_jordan_link_rates}
\end{equation}
Thus, the relevant nearest-layer rates cannot grow parametrically down the Jordan chain. Chirality also implies that links already present in a shorter chain are inherited unchanged when additional dark-dimers are appended upstream.

Next, noting that:
\begin{equation}
    S_M^{M-1}=\left(\prod_{r=1}^{M-1}(S_M)_{r,r+1}\right) \ketbra{u_1}{u_M},
\end{equation}
it is therefore natural to define the effective rate entering the longest-chain term by,
\begin{equation}
    g_{\mathrm{eff},M} :=\left(\prod_{r=1}^{M-1} g_r \right)^{\frac{1}{M-1}}= \left\|S_M^{M-1}\right\|_{2 \rightarrow 2}^{\frac{1}{M-1}},
    \quad 0<g_{\mathrm{eff},M}\le\gamma_R.
    \label{eq:geff_exact_definition}
\end{equation}
 
In the late-time Jordan tail, the term carrying the largest power of $t$ is the one that contains the full chain length and therefore controls the strongest possible $N$-dependent delay. Eq.~\eqref{eq:geff_exact_definition} makes its coefficient exact:
\begin{equation}
    e^{-\Delta t}\frac{t^{M-1}}{(M-1)!}\left\|S_M^{M-1}\right\|_{2 \rightarrow 2} = e^{-\Delta t}\frac{(g_{\mathrm{eff},M}t)^{M-1}}{(M-1)!}.
\end{equation}
We therefore approximate the preparation-time crossing by:
\begin{equation}
    e^{-\Delta T_\epsilon}\frac{(g_{\mathrm{eff},M}T_\epsilon)^{M-1}}{(M-1)!}\sim\epsilon.
    \label{eq:highest_jordan_ansatz_crossing}
\end{equation}
This is the core relation of our argument. Using the $W_{-1}$ branch of the Lambert function, we obtain:
\begin{equation}
    T_\epsilon^{(ans)} = -\frac{M-1}{\Delta} W_{-1} \left[ -\frac{\Delta}{g_{\mathrm{eff},M}}\, \frac{((M-1)!\epsilon)^{1/(M-1)}}{M-1}
    \right].
\end{equation}

Next, since Eq.~\eqref{eq:geff_exact_definition} gives $0<g_{\mathrm{eff},M}\le\gamma_R$, we note that:
\begin{equation}
    e^{-\Delta t}\frac{(g_{\mathrm{eff},M}t)^{M-1}}{(M-1)!} \le e^{-\Delta t}\frac{(\gamma_R t)^{M-1}}{(M-1)!},
    \label{eq:highest_jordan_gamma_envelope}
\end{equation}
and so we can define the upper-bound:
\begin{equation}
    T_\epsilon^{(env)} = -\frac{M-1}{\Delta} W_{-1} \left[ -\frac{1}{2}\frac{((M-1)!\epsilon)^{1/(M-1)}}{M-1} \right],
    \label{eq:gamma_envelope_lambert}
\end{equation}
where $T_\epsilon^{(env)} \geq T_\epsilon^{(ans)}$ whenever the late-time $W_{-1}$ branch is present. Using the Robbins-Stirling form,
\begin{equation}
    n!=\sqrt{2\pi n}\,n^n e^{-n}e^{r_n},
    \quad \frac{1}{12n+1}<r_n<\frac{1}{12n},
\end{equation}
the argument of $W_{-1}$ becomes:
\begin{equation}
    -\frac{1}{2e} \exp\left[\frac{\ln \left(\epsilon\sqrt{2\pi (M-1)} \right)+r_{M-1}}{M-1}\right] = -e^{-1-a},
    \label{eq:dimensionless_lambert_argument}
\end{equation}
where $a = \frac{L'-r_{M-1}}{M-1} + \ln 2$ and $L' = -\ln [\epsilon\sqrt{2\pi (M-1)}]$.

If $a>0$, we can apply Chatzigeorgiou's bound \cite{Chatzigeorgiou2013}:
\begin{equation}
    -1-\sqrt{2a}-a < W_{-1} [-e^{-1-a}] < -1-\sqrt{2a} - \frac{2}{3}a,
\end{equation}
to obtain the upper bound:
\begin{equation}
    T_\epsilon^{(env)} < \frac{M-1}{\Delta}(1+\sqrt{2a} + a)
\end{equation}
which scales as $\mathcal{O}(M/\Delta) \sim \mathcal{O}(N/\gamma_R)$. We can see this clearly by checking the scaling of $a$:
\begin{equation}
    a=  -\frac{\ln \epsilon}{M-1} -\frac{\ln [2\pi (M-1)]}{2(M-1)}  - \frac{r_{M-1}}{M-1} + \ln 2,
\end{equation}
and noting that all terms scale either $\mathcal{O}(1)$ or $\mathcal{O}(M^{-1})$. Note that the condition $a>0$ is fulfilled by any fixed $0<\epsilon<1$, so there are in fact no restrictions on the usage of the bound. Thus, we have established linear-in-$N$ scaling of preparation-time to steady state.

\bibliography{Ref}

\end{document}